\documentclass[iop,apj]{emulateapj}
\usepackage{amsmath,amssymb,amstext,float}

\usepackage[citecolor=black,linkcolor=black]{hyperref}
\usepackage{color}

\usepackage[all]{hypcap} 
\usepackage{aas_macros}
\usepackage{natbib}
\usepackage{mathtools}
\shorttitle{Statistics Of Driven Reconnection In The Corona}
\shortauthors{Knizhnik et al.}
\newcommand{\beg}[1]{\begin{equation}\label{#1}}
\newcommand{\done}{\end{equation}}
\newcommand{\pd}[2]{\frac{\partial #1}{\partial #2}}

\newcommand{\vecB}{\textbf{B}}

\newcommand{\vecS}{\textbf{S}}

\newcommand{\vecv}{\textbf{v}}

\newcommand{\curl}[1]{\nabla\times{#1}}
\newcommand{\divv}[1]{\nabla\cdot{#1}}
\numberwithin{equation}{section}
\begin{document}

\title{Power-law Statistics Of Driven Reconnection In The Magnetically Closed Corona}
\author{K. J. Knizhnik\altaffilmark{1}, V.M. Uritsky\altaffilmark{2,3}, J.A. Klimchuk\altaffilmark{2}, and C. R. DeVore\altaffilmark{2}}
\altaffiltext{1}{National Research Council Research Associate Residing At Naval Research Laboratory, 4555 Overlook Ave SW, Washington, DC 20375, USA}
\altaffiltext{2}{Heliophysics Science Division, NASA Goddard Space Flight Center, 8800 
Greenbelt Rd, Greenbelt MD 20771}
\altaffiltext{3}{Catholic University of America, 620 Michigan Ave NE, Washington, DC 20064}

\begin{abstract}
Numerous observations have revealed that power-law distributions are ubiquitous in energetic solar processes. Hard X-rays, soft X-rays, extreme ultraviolet radiation, and radio waves all display power-law frequency distributions. Since magnetic reconnection is the driving mechanism for many energetic solar phenomena, it is likely that reconnection events themselves display such power-law distributions. In this work, we perform numerical simulations of the solar corona driven by simple convective motions at the photospheric level. Using temperature changes, current distributions, and Poynting fluxes as proxies for heating, we demonstrate that energetic events occurring in our simulation display power-law frequency distributions, with slopes in good agreement with observations. 
We suggest that the braiding-associated reconnection in the corona can be understood in terms of a self-organized criticality model driven by convective rotational motions similar to those observed at the photosphere.
\end{abstract}

\maketitle

\section{introduction}\label{sec:intro}
It is well known that many solar phenomena display power-law frequency distributions. Starting with observations of power laws in hard X-ray bursts \citep{Datlowe74, Dennis85}, numerous other such power laws have been observed. Solar flare hard X-rays, produced by nonthermal bremsstrahlung from accelerated coronal electrons traveling downwards toward the chomosphere, soft X-ray emission, produced by evaporated chromospheric plasma, and Extreme Ultraviolet (EUV) radiation display power-law frequency distributions in peak flux, energy, and duration \citep[][and references therein]{Aschwanden16}. Power-law slopes have also been measured in the distributions of solar energetic particle event fluxes \citep{Gabriel96}, radio bursts \citep{Iwai13, Das97, Ning07}, and Ellerman bombs \citep{Georgoulis02}. \par
In the context of energetic events such as solar flares, magnetic reconnection plays a fundamental role, converting magnetic energy into thermal energy and accelerating particles \citep{Knizhnik11,Dahlin17} and (at least indirectly) causing many of the signatures described above. A key question, therefore, is: do reconnection events in the solar corona occur with power-law frequency distributions? \par
Recently, we used numerical simulations to study the role of magnetic reconnection in structuring and heating the solar atmosphere \citep{Knizhnik15, Knizhnik17}. Our simulations assumed a plane-parallel \citet{Parker72} corona between two plates, driven by random footpoint motions to inject magnetic energy and helicity into the solar atmosphere. We established that magnetic reconnection transported helicity and energy throughout the solar atmosphere, where it accumulated at flux-system boundaries to form filament channels, in a process known as magnetic helicity condensation \citep{Antiochos13}. A large range of scales is involved in the process, from the granule-sized helicity-injection scale up to the filaments-sized helicity-accumulation scale. Hence, this reconnection driven inverse-cascade can be used to investigate the statistics of its reconnection events. By measuring their size, energy, and temporal distribution, we quantify the statistics of these events in our model corona. \par
Previous studies of the statistics of reconnection events in Parker coronae \citep[e.g.,][]{Krasnoselskikh02, Morales08a, Morales08b, Morales09, LF15, LF16} were primarily cellular automaton models. In these models, driving and instability criteria were applied to the system, prescribed according to a set of discrete rules. For example, in the model of \citet{Morales08a}, the driving mechanism was a random horizontal (integer) displacement of the footpoints of magnetic strands, and the instability criterion was a critical angle between adjacent strands. These studies yielded power-law distributions for many measured quantities, with spectral indices in the observed range.  \par
In this study, we calculate the frequency distribution of magnetic reconnection events occurring self-consistently in our numerical model of helicity condensation in a randomly driven \citet{Parker72} corona. This situation is relevant to the Sun, where photospheric convection twists and braids the coronal magnetic field, causing current sheets to develop and reconnect. Since the nature of the driving is a key difference between our model and the cellular automata described above, and is also a key ingredient in modeling the solar photosphere, we discuss briefly the manner in which the coronal magnetic field is driven at the photospheric level. \par
Although the photospheric flows that inject energy into the corona are highly complex \citep{Schmieder14} from the standpoint of energy injection, the important flows are those that have significant vorticity, twisting the field internal to individual flux tubes or braiding flux tubes around one another \citep{Longcope07}. On the other hand, compression of the field by radial outflows with negligible vorticity, the dominant form of convection \citep{Schmieder14}, is not expected to cause significant reconnection, and therefore energy release, in the magnetic field. Instead, these radial outflows act to concentrate the field in narrow intergranular lanes \citep{Nordlund09}. In these regions, vortical flows have been observed at the photosphere on the scales of both granules \citep[e.g.][]{Bonet08,Bonet10,VD11,VD15} and supergranules \citep{Brandt88,Attie09,Langfellner15}, with mean lifetimes of about a day \citep{Rieutord10}. Therefore, it suffices to model the energy injection into the corona with simple twisting motions localized in narrow annuli, as has been done by many authors \citep[e.g.][]{WilmotSmith10, Rappazzo13}. Furthermore, observations strongly imply a preferred sign for this vorticity in each hemisphere of between 55-80\% \citep{Martin94, Pevtsov03}. Indeed, simulations have found that the amount of observed coronal structure is consistent with a $\sim 75\%/25\%$ negative/positive helicity injection preference in the northern hemisphere \citep{Knizhnik15,Knizhnik17,Knizhnik17b}.\par


The key point, however, is that our simulation is a driven $3$D model solar corona that self-consistently undergoes energy releasing events via numerical reconnection, rather than a cellular automaton that follows a set of prescribed rules to determine the driving and/or instability criterion. The fact that reconnection occurring in the corona allows individual magnetic field lines to relax means that the energy injected into the corona depends critically on the dynamics in the corona itself, rather than simply on the photospheric motions \citep{Klimchuk06, Klimchuk15}. Nevertheless, the photospheric driving plays a critical role in the energetics of the corona, and determining how reconnection statistics depend on the driving is an important question. Using our model we will demonstrate that multiple reconnection events in the driven magnetically-closed corona obey broad-band power-law distributions similar to those found in coronal observations, and show that the measured slopes are quantitatively consistent with both cellular automaton models and with observations.\par
This paper is organized as follows: in \S \ref{sec:model} we describe our numerical simulations, including how we implement a driver that closely approximates the relevant features of photospheric convection. In \S \ref{sec:Results} we describe the results of our simulations, before interpreting them in the context of the solar corona in \S \ref{sec:implications}.
\section{Numerical Model}\label{sec:model}
Our simulations solve the equations of magnetohydrodynamics (MHD) using the Adaptively Refined Magnetohydrodynamics Solver \citep[ARMS;][]{DeVore08} in three Cartesian dimensions. The equations have the form
\beg{cont}
\pd{\rho}{t}+\divv{\rho\vecv}=0,
\done
\beg{momentum}
\pd{\rho\vecv}{t} + \divv{\left( \rho\vecv\vecv \right)} = - \nabla P + \frac{1}{4\pi} \left( \curl{\vecB} \right) \times \vecB,
\done
\beg{energy}
\pd{U}{t}+\divv{\Big\{\left(U+P+\frac{B^2}{4\pi}\right)\vecv-\frac{\vecB(\vecv\cdot\vecB)}{4\pi}\Big\}}=0.
\done
\beg{induction}
\pd{\vecB}{t} = \curl{ \left( \vecv \times \vecB \right)}.
\done
where
\beg{totenergydensity}
 U=\epsilon+K+W
\done
is the total energy density, the sum of the internal energy density
\beg{internal}
\epsilon=\frac{P}{\gamma-1},
\done
kinetic energy density
\beg{kinetic}
K=\frac{\rho v^2}{2},
\done
and magnetic energy density
\beg{magneticenergydensity}
W=\frac{B^2}{8\pi}.
\done
In these equations, $\rho$ is mass density, $T$ is temperature, $P$ is thermal pressure, $\gamma$ is the ratio of specific heats, $\vecv$ is velocity, $\vecB$ is magnetic field, and $t$ is time. We close the equations via the ideal gas equation,
\beg{ideal}
P = \rho RT,
\done
where $R$ is the gas constant.\par

ARMS uses finite-volume representations of the variables to solve the system of equations. Its Flux Corrected Transport algorithms \citep{DeVore91} provide minimal, though finite, numerical dissipation, which allows reconnection to occur. As a result, to a very good approximation, ARMS conserves the magnetic helicity in the system \citep{Knizhnik15, Knizhnik17}.\par

We set up a model coronal field that is initially straight and uniform between two plates as in our previous work \citep{Knizhnik15, Knizhnik17, Knizhnik17b}. 
In this model, straight flux tubes represent coronal loops whose apex is located in the center of the domain, with the two plates representing the photosphere. Our domain size is $[0,L_x] \times [-L_y,L_y] \times [-L_z,L_z]$, where $x$ is taken normal to the photosphere (the vertical direction) and we set $L_x=1$, and $L_y=L_z=2.5$. At all six boundaries, we use zero-gradient conditions, and the four side walls have open boundary conditions. Closed boundary conditions are employed at the top and bottom, where the magnetic field is line tied at the high-$\beta$ photosphere. The footpoints of the field lines do not move in response to magnetic forces, but do respond to imposed boundary flows to mimic driving at the plasma-dominated photosphere.\par

We set the initial, uniform values in our dimensionless simulations to $\rho_0=1$, $R=0.05$, $P_0=0.1$, and $B_0=\sqrt{4\pi}$. These choices set the temperature, $T=2$, the Alfv\'en speed, $c_{A0}=B_0 / \sqrt{4\pi\rho_0} = 1$, and the plasma beta, $\beta_0=8\pi P_0/B_0^2=0.2$. $\beta\ll1$ corresponds to a magnetically dominated plasma, which is generally true of the corona. The results discussed below will be given in simulation time, which is normalized to the time required for an Alfv\'en wave at unit speed ($c_{A0}=1$) to travel the distance separating the top and bottom plates ($L_x=1$).\par

A key difference between this study and our previous work \citep{Knizhnik15,Knizhnik17}, is that in this study energy conservation is ensured via Eq. (\ref{energy}). In our previous simulations, the adiabatic temperature equation was solved, and any changes in the internal energy density were due to compression or expansion. By solving the conservative form of the energy equation, we guarantee that magnetic reconnection can convert magnetic energy into both plasma heating and bulk motions. Since there is no conduction and/or radiation in our simulations, any increases in the temperature are expected to continually accumulate. Nevertheless, energy conservation guarantees that we accurately capture the conversion of magnetic energy into plasma energy during magnetic reconnection. \par

\begin{figure*}
\centering\includegraphics[scale=0.35,trim=0.0cm 0cm 0cm 0cm, clip=true]{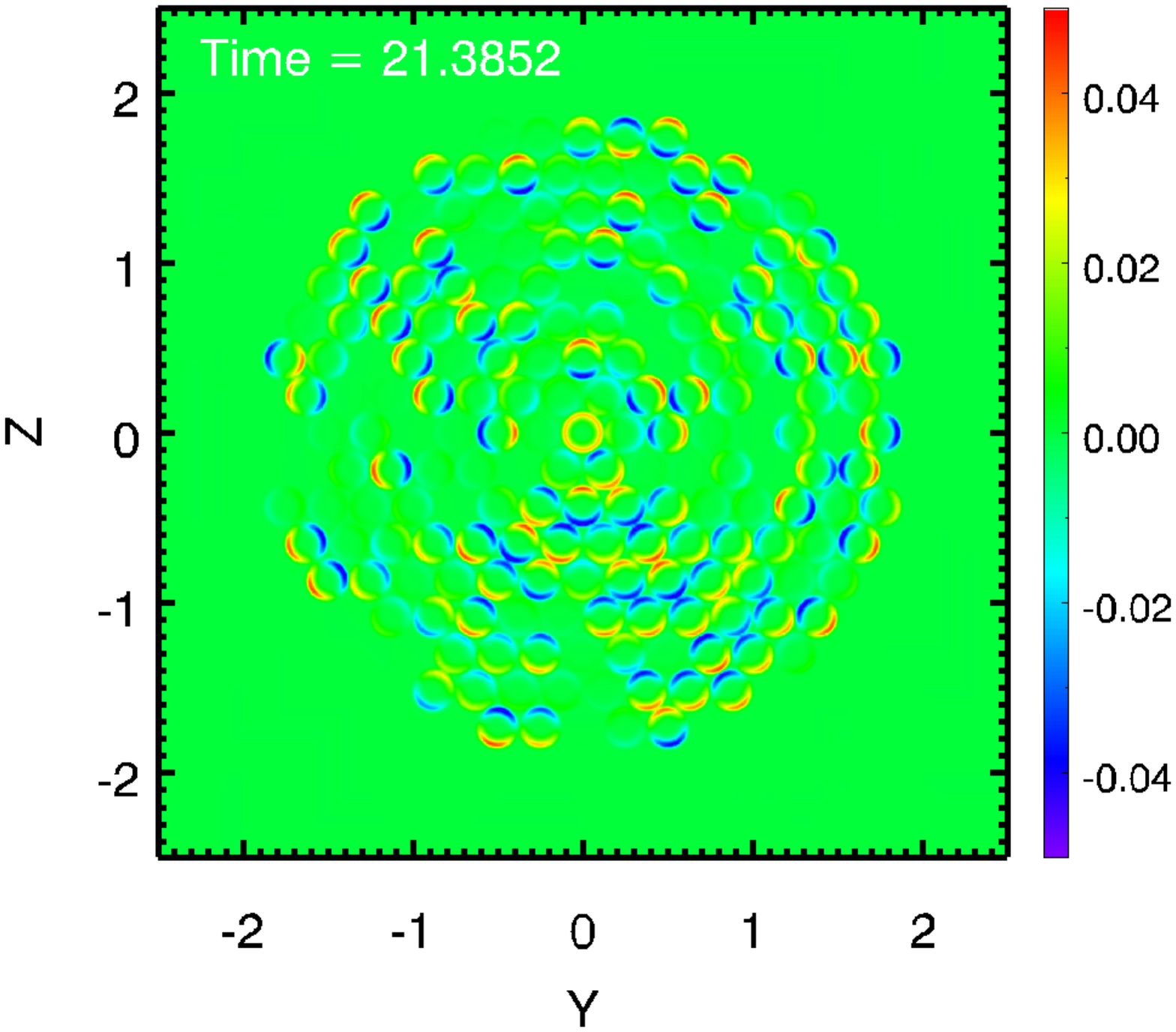}
\centering\includegraphics[scale=0.35,trim=0.0cm 0cm 0cm 0cm, clip=true]{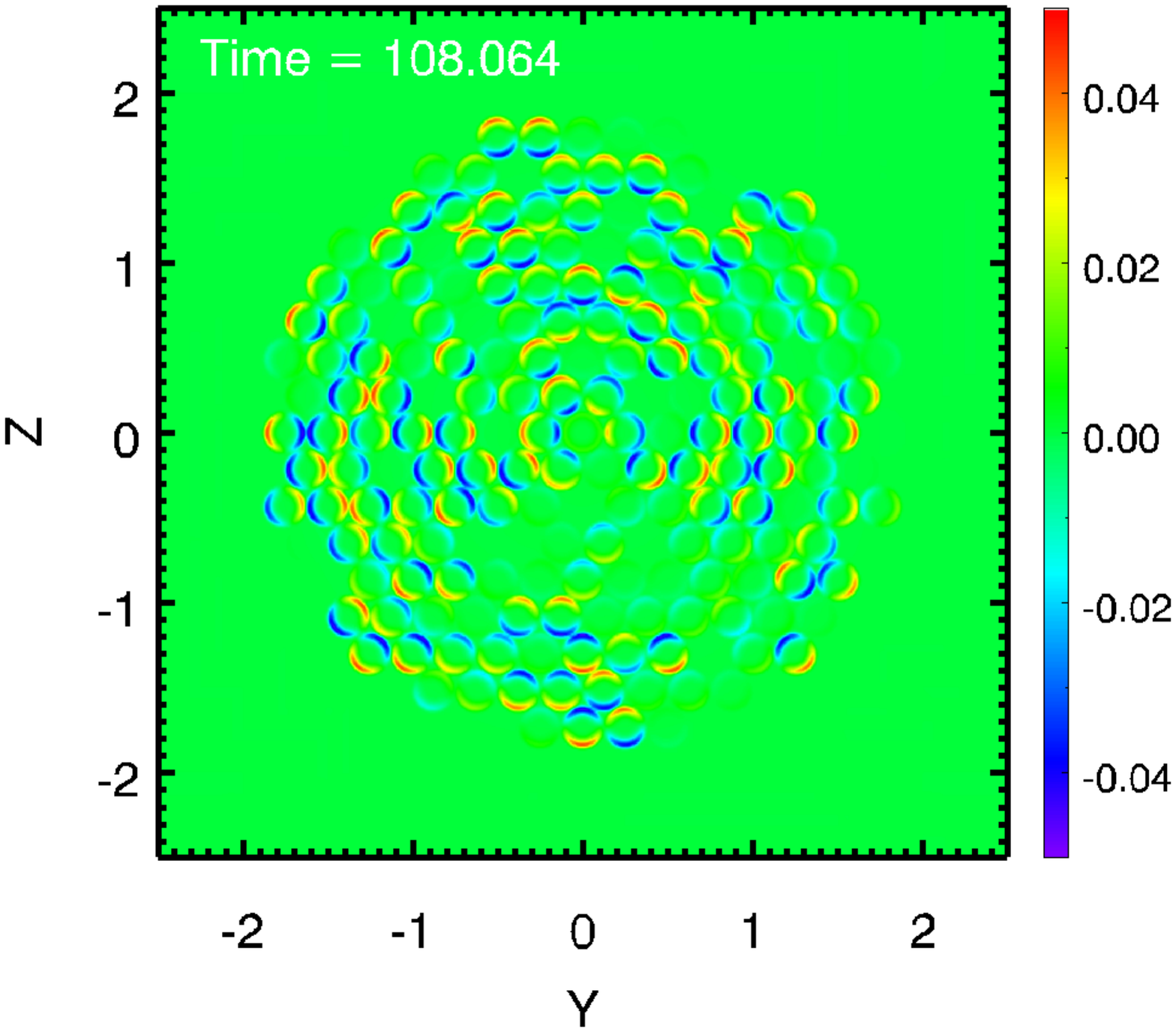}
\centering\includegraphics[scale=0.35,trim=0.0cm 0.5cm 0cm 0cm, clip=true]{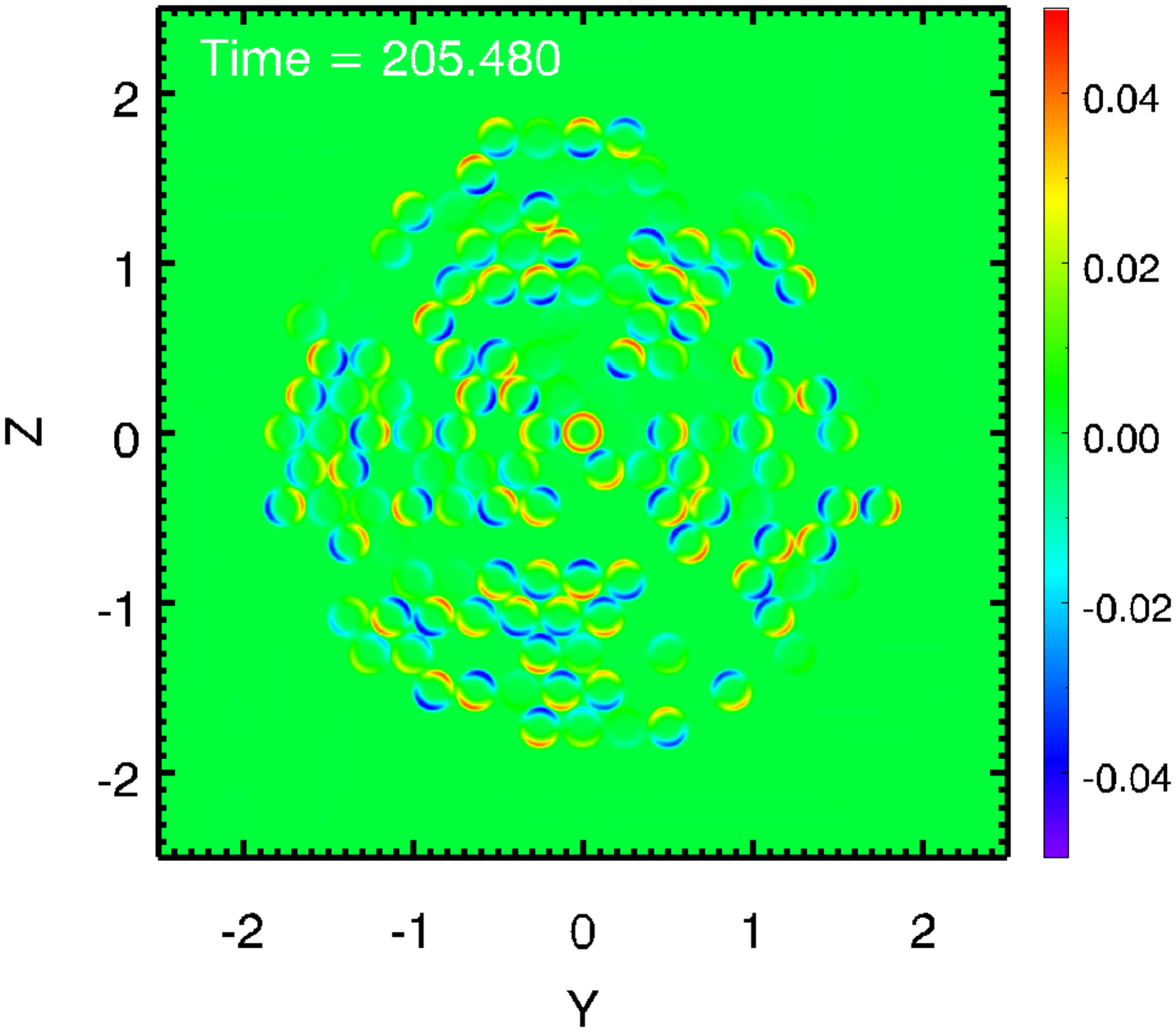}
\vspace{20mm}\caption{$V_\phi$ on the bottom plate near the beginning, middle, and end of the simulation. Red/yellow (blue/teal) represent counter-clockwise/positive (clockwise/negative) velocity.}
\label{fig:vphi}
\end{figure*}

We perform our simulations with a photospheric velocity profile that contains $199$ convection cells on both the top and bottom plates. Each twist cycle for each cell consists of a slow ramp-up phase, followed by a slow decline phase with the angular velocity of each individual cell having the form:
\beg{Omega}
\Omega(r,t) = A g(r)f(t)
\done
with $A$ the flow amplitude, $g(r)$ its spatial profile, and $f(t)$ its temporal profile. 
Fig. \ref{fig:vphi} shows the azimuthal velocity on the bottom plate at three different times during the simulation, where the azimuthal angle is defined with respect to the center of the plane, not the axis of each rotational cell.
The exact forms and value of $A$ are specified by attempting to maximize the realism of our simulations and reproducing the following salient observational features of photospheric convection:
1) helicity injection preference,
2) temporal profile, and
3) spatial profile.
Each of these will be discussed below.

\subsection{Helicity Injection Preference}
Each cell twists up the coronal magnetic field, with $\sim 75\%/25\%$ of the cells injecting positive/negative magnetic helicity. The sense of rotation of each cell, and hence the sign of helicity it injects, is assigned randomly, by giving each cell on the bottom plate a random number $f \in [0,1]$, and letting
\beg{A}
A =
\begin{cases}
-\Omega_0 & \text{if } f>0.25\\
+\Omega_0 & \text{if } f<0.25
\end{cases}
\done
where we choose $\Omega_0=1.875$. The sense of rotation of cells on the top must necessarily be opposite to that of the cells on the bottom for a pair of cells to inject a net magnetic helicity. This procedure is exactly the same as that detailed in \citet{Knizhnik17}. Those simulations were shown to conserve helicity, an important requirement to achieving power-law statistics over many size scales \citep{Chou01}. As a result, photospheric motions inject $\sim 75\%/25\%$ positive/negative magnetic helicity, in approximate agreement with observations \citep[e.g.,][]{Pevtsov03}. \par 

\subsection{Temporal Profile}
We specify the temporal profile of our cells to have the form:
\beg{foft}
f(t) =
\begin{cases}
\frac{1}{2}\Big[1-\cos\Big(2\pi\frac{t-t_s}{\tau}\Big)\Big] & \text{if } t>t_s\\
0 & \text{if } t<t_s
\end{cases}
\done
Here, $t_s$ is the start time of the cell, so that it only turns on at $t=t_s$, whereupon it follows this oscillating temporal profile. We set $\tau=13.4$ to be the period during which the maximum angle of the clockwise rotation within each cell is $\phi_{max} = \pi$. Since the twisting is occurring on both the top and bottom plates, the maximum rotation angle of each flux tube is $2\pi$. To randomize the driving, we stagger the phase of each of the convection cells by assigning each top/bottom pair a random number between $0$ and $\tau$, so that each cell turns on at a slightly different time. Again, this is done to maximize the realism of our simulation, since convection on the photospheric level is not coherent, but is constantly randomly appearing and disappearing \citep{Rieutord10}.  At each of the three times
shown in Fig. \ref{fig:vphi},
a different number of convection cells are turned on, having been staggered randomly in their individual phases. In addition, the helicity injection preference is evident, with approximately three quarters of the cells rotating in one sense, and the remaining cells rotating in the other. The sense of rotation of a given cell remains fixed throughout the simulation, however.
\par

\subsection{Spatial Profile}
\begin{figure*}
\centering\includegraphics[scale=0.75, trim=0.0cm 0.0cm 0.0cm 0.0cm]{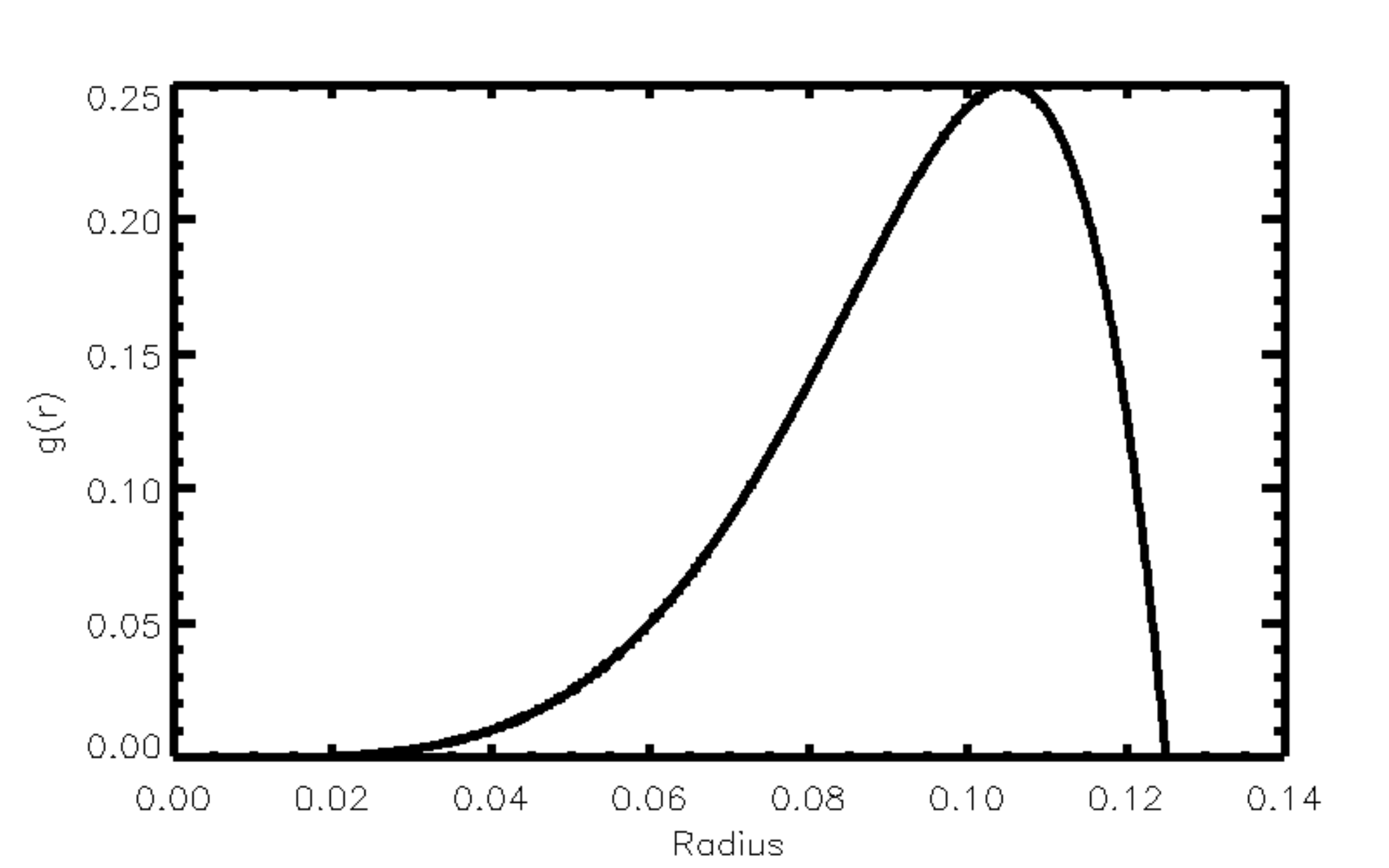}
\caption{g(r) for a single convection cell, showing that the velocity profile is concentrated in the outer annulus.}
\label{fig:gofr}
\end{figure*}

We specify the spatial profile of our cells to have the form:
\beg{gofr}
g(r) = \Big(\frac{r}{a_0}\Big)^4 - \Big(\frac{r}{a_0}\Big)^8,
\done
\newline
shown in Fig. \ref{fig:gofr}. We set $a_0=0.125$ to be the flow radius. This form of the flow velocity is incompressible and satisfies $v_n = v_x = 0$. Such a profile $g(r)$ creates a strong concentration of the twisting profile in the outer annulus of the convection cell. In addition, the convective cells are arranged compactly in a hexagonal pattern, permitting them to be placed as closely as possible to each other. This concentration of helicity injection in narrow lanes between closely packed convection cells is crucial for driven heating models since on the Sun, energy, in the form of twist, is injected in narrow intergranular lanes which occur as a result of converging radial outflows from adjacent supergranules \citep{Nordlund09}. \par
Our numerical resolution supports 32 grid points across the diameter of each one of our convection cells. We impose $15$ complete rotations of each top/bottom cell pair, followed by a brief relaxation phase in which no twisting is applied. This extended driving period allows any initial transient behavior to die off, and the system has time to forget its initial untwisted state.

\section{Results}\label{sec:Results}
The photospheric motions we impose inject energy into the coronal field in the form of magnetic energy, which gets converted primarily into heating as a result of magnetic reconnection. In 
the top panel of 
Fig. \ref{fig:energies} we plot the instantaneous and time integrated 
normal component of the Poynting flux at the photospheric boundary
 over the course of the simulation. 
These are given by, respectively, 
\beg{PF}
F(t) = \int_S{S_n(t) dS},
\done
and
\beg{tip}
\mathcal{F}(t) = \int_0^t{F(t') dt'}. 
\done
Here
\beg{Sn}
S_n = \frac{1}{4\pi}\Big[(\vecB_h\cdot\vecB_h)\vecv_n - (\vecv_n\cdot\vecB_h)\vecB_h\Big] ,
\done
with $v_n$ the normal component of the velocity (which vanishes for our flow profile), $\vecv_h$ is the horizontal velocity, and $\vecB_h$ is the horizontal magnetic field.
The instantaneous Poynting flux shows that the system takes some time to reach a quasi-steady state. After an initial steep increase and subsequent drop due to the relaxation of magnetic field lines after reconnection, the Poynting flux reaches a statistical equilibrium around t=50. Interestingly, there is clearly a signature of the driving, with oscillations of period $\sim \tau$ appearing, despite the convective cells being out of phase with each other. The reason is that the relative phasing of the cells is fixed for the duration of the simulation. Due to the random start times of the cells, the average driving across the entire system has a small variability, and it repeats with the same period as the individual cells. 
In the bottom panel of Fig. \ref{fig:energies}, we plot the magnetic, internal, and kinetic energy as a function of time in our simulation. Note that the kinetic energy has been multiplied by a factor of $10^3$ to make it visible on the plot. Since there are no energy sinks in our model, the plasma energy steadily rises. Nevertheless, the kinetic energy is more or less steady and the magnetic energy settles to a very low rate of increase by about $t=50$, likely due to the constant increase in plasma energy - and, therefore, plasma $\beta$. The energy conversion rate in our simulation is determined from energy balance considerations. The energy released by reconnection events between two times $t_1$ and $t_2$ is given by
\beg{ERx}
E_{Rx} = \int_{t_1}^{t_2}{\Big(\frac{dU}{dt} - \frac{dW}{dt}\Big)dt}
\done
The average energy conversion rate is, therefore,
\beg{PRx}
\langle P_{Rx} \rangle = \langle \frac{dU}{dt}\rangle - \langle \frac{dW}{dt} \rangle .
\done
Energy conservation implies that
\beg{energy_cons}
\langle \frac{dU}{dt}\rangle = F(t),
\done
so the average energy conversion rate is 
\beg{Prxfinal}
\langle P_{Rx} \rangle = F(t) - \langle \frac{dW}{dt} \rangle
\done
This equation states that the energy that powers reconnection is given by the difference between the energy injected into the system and the change in magnetic energy. We plot $\langle P_{Rx} \rangle$ in the bottom panel of Fig. \ref{fig:energies}. The peaks in the reconnection power are clearly correlated with the peaks in kinetic energy, in agreement with the intuition that reconnection generates strong outflows from the reconnection site. The roughly constant amplitude and frequency of the oscillations for times $t>50$ suggests that the energy conversion is roughly stationary.
\par

\begin{figure*}
\centering\includegraphics[scale=0.25,trim=0.0cm 0.0cm 0cm 0.0cm, clip=true]{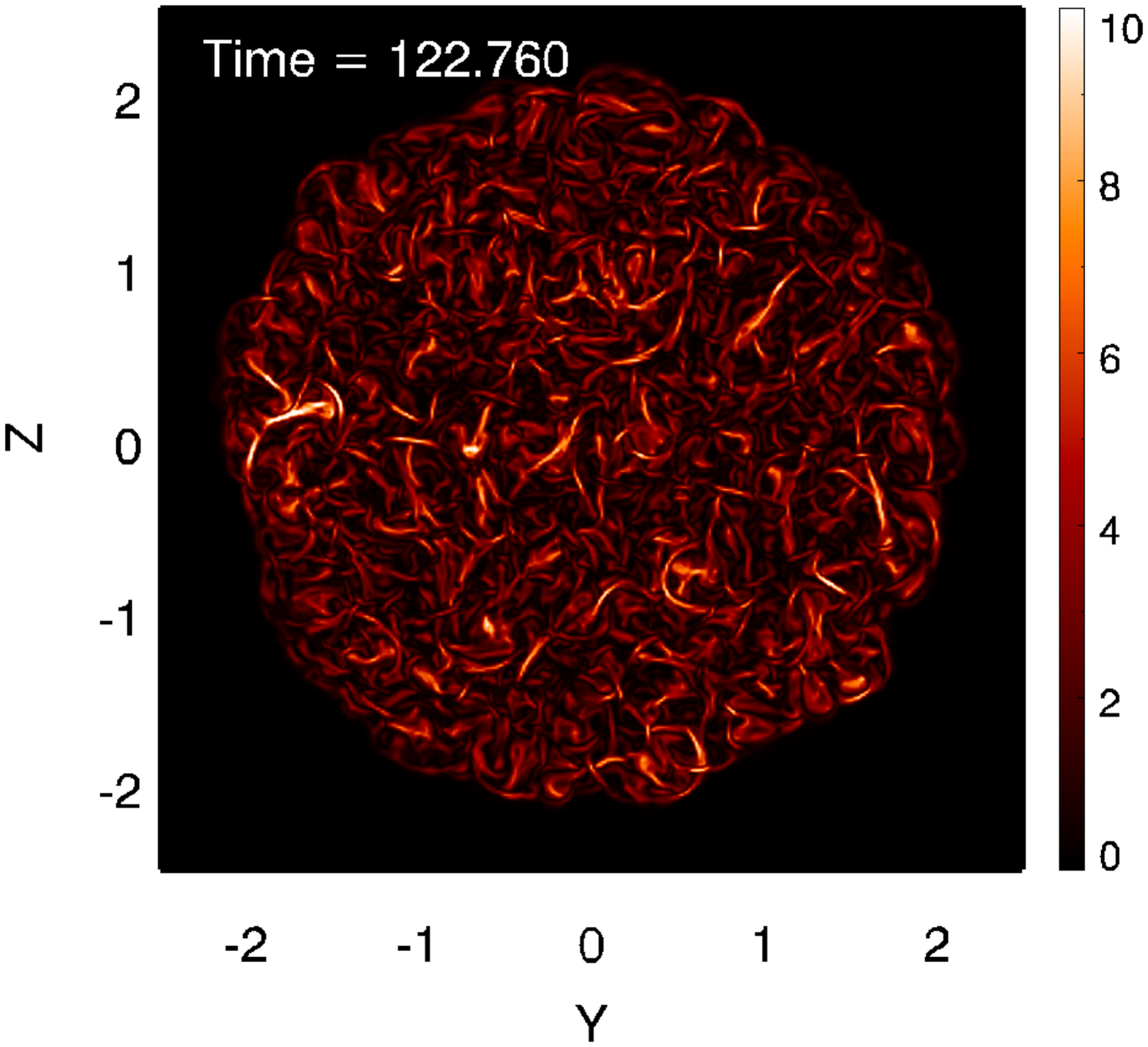}
\centering\includegraphics[scale=0.25,trim=0.0cm 0.0cm 0cm 0.0cm, clip=true]{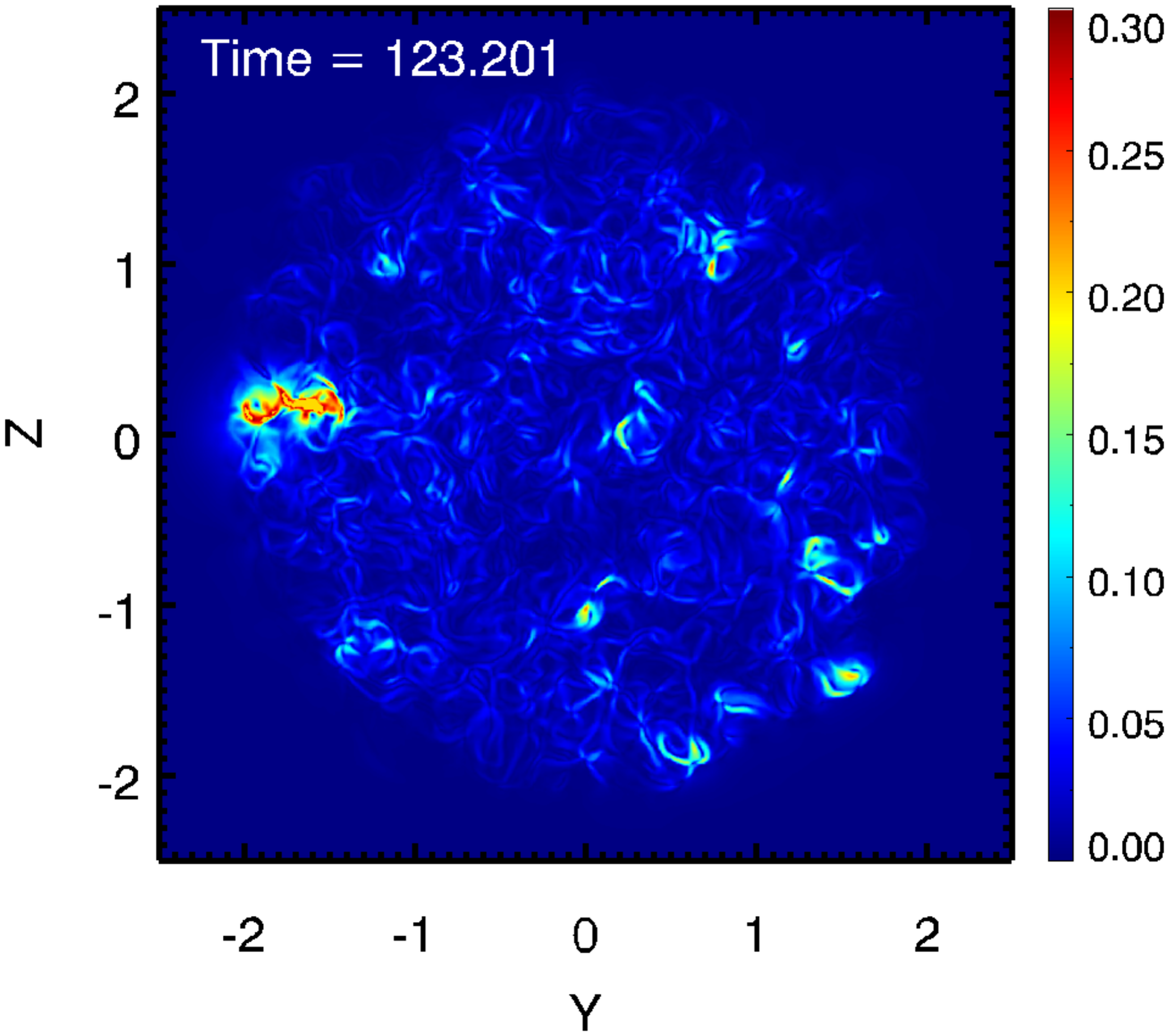}
\centering\includegraphics[scale=0.25,trim=0.0cm 0.0cm 0cm 0.0cm, clip=true]{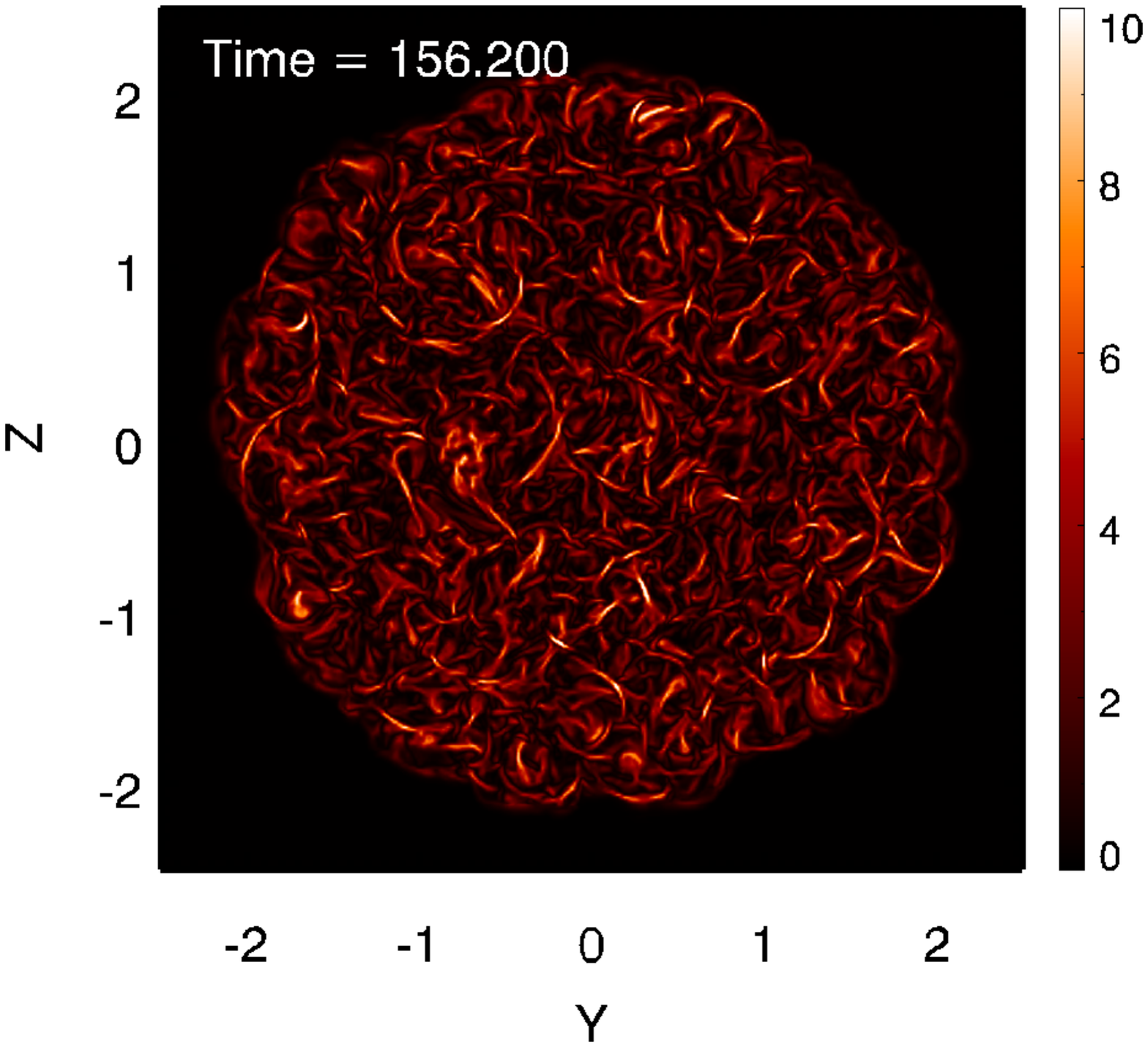}
\centering\includegraphics[scale=0.25,trim=0.0cm 0.0cm 0cm 0.0cm, clip=true]{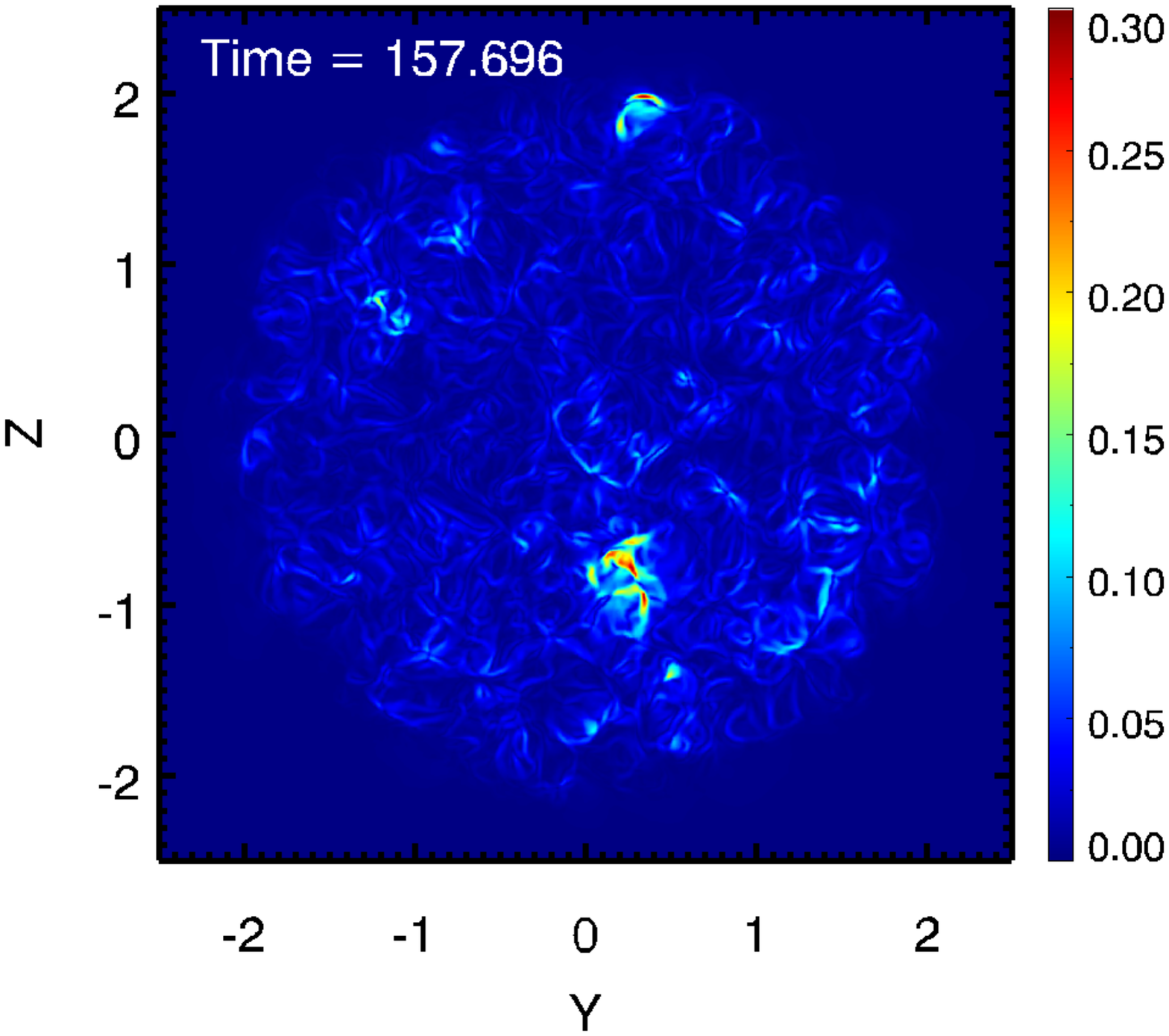}
\centering\includegraphics[scale=0.25,trim=0.0cm 0.0cm 0cm 0.0cm, clip=true]{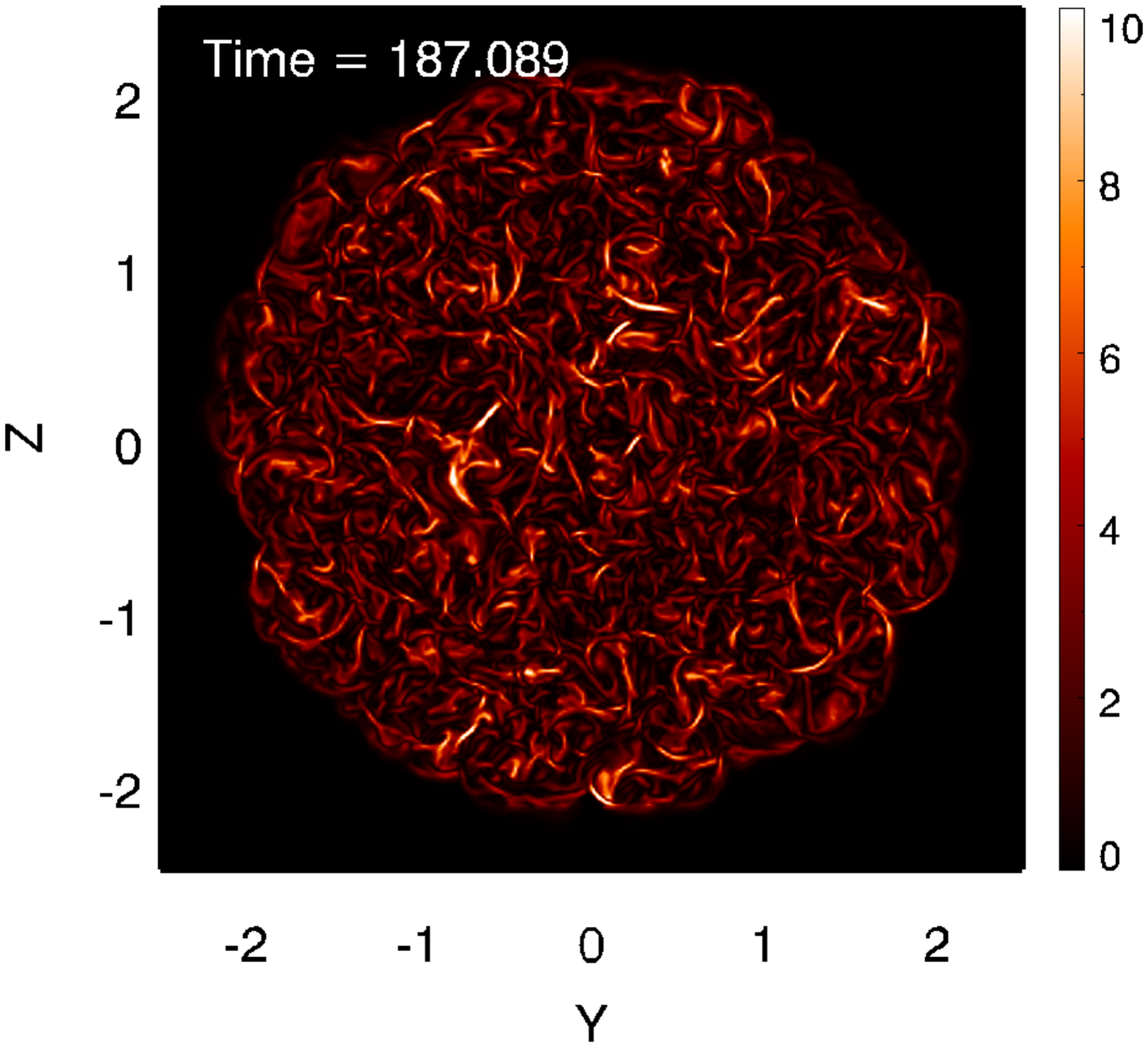}
\centering\includegraphics[scale=0.25,trim=0.0cm 0.0cm 0cm 0.0cm, clip=true]{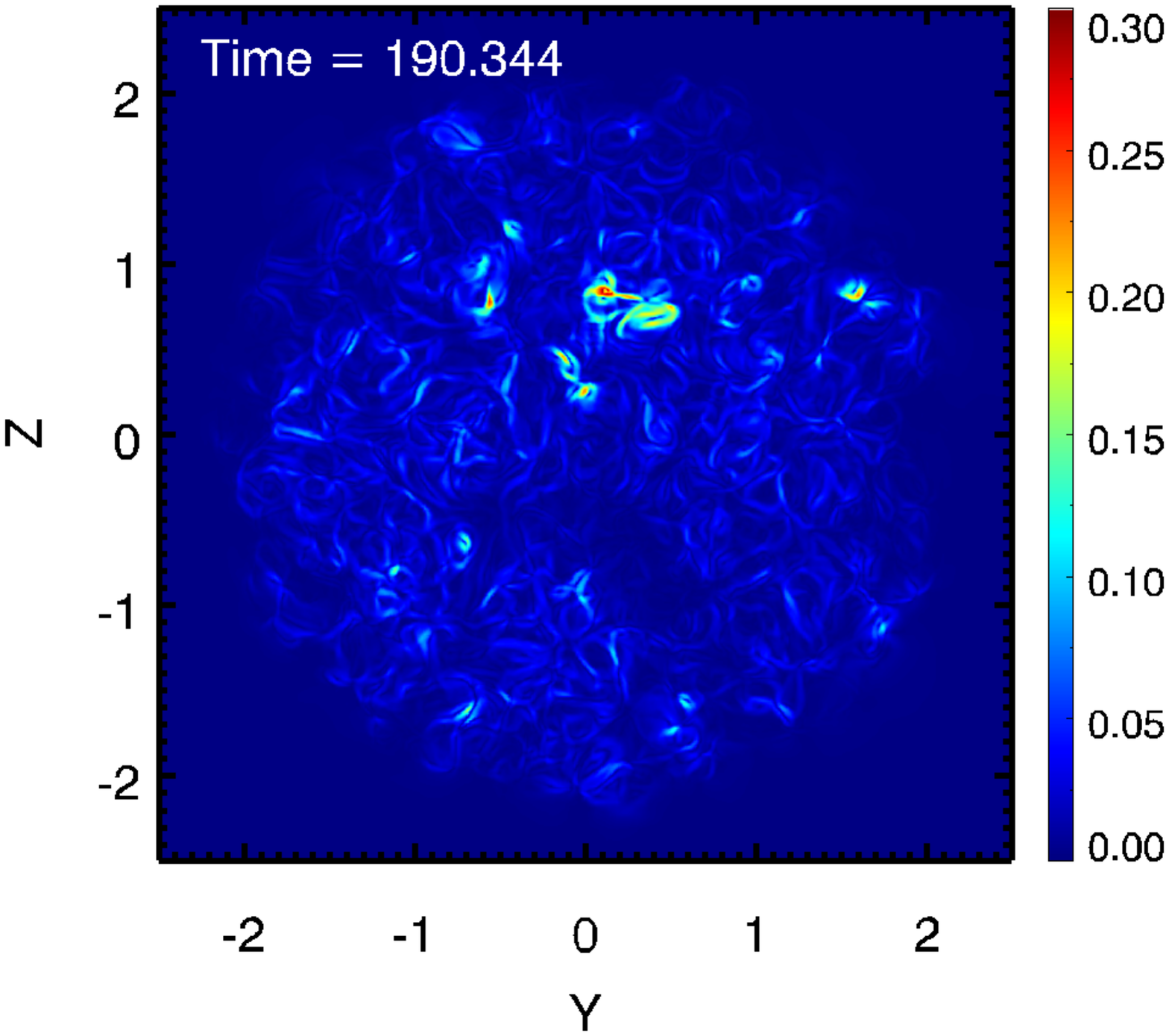}
\vspace{0mm}\caption{Left: Current density in the midplane at three different times during the simulation. Right: $S_h$ at times shortly after the current density frames. }
\label{fig:JandS}
\end{figure*}

The stress injected by the surface motions creates numerous thin current sheets in the simulation domain. In the left panels of Fig. \ref{fig:JandS}, we show some representative frames of the current density magnitude  in the midplane at different times during the simulation. These frames show the development of thin current sheets throughout the twisting region. 
These current structures appear at various different locations over the 
course of the simulation and have a wide range of transverse scales. 
Both long and short, weak and strong, thin current sheets form and 
dissipate throughout the duration of the driving. The transverse scale 
of these current sheets is, in many cases, greater than the length scale 
of the driving flows. Furthermore, the dissipation of many of the strong, extended current sheets causes a significant interaction with adjacent structures which, in turn, affect their own nearest neighbors. The continual driving of the system prevents currents from dissipating entirely, however, with new current sheets constantly being formed as old ones are dissipated.\par
The constant dissipation of these current sheets contributes significantly to the heating of the coronal plasma. One way to understand where the heating occurs is to calculate the horizontal Poynting flux density at the midplane. In ideal MHD, the Poynting flux density is given by
\beg{Sdens}
\vecS = -\frac{1}{4\pi} (\vecv \times \vecB) \times \vecB .
\done
Let $S_h$ be the magnitude of the horizontal (y,z) component of $\vecS$:
\beg{Sh}
S_h = \frac{1}{4\pi}|(\vecB\cdot\vecB)\vecv_h - (\vecv\cdot\vecB)\vecB_h|
\done
Regions of locally enhanced $S_h$ are associated with Alfv\' enic jets formed as a result of reconnection, and the relaxation of magnetic field lines following reconnection. These will be sites of intense magnetic energy release and conversion. Of course, these processes are not necessarily constrained to occur in a two-dimensional plane, but because the reconnecting component of the magnetic field is largely the in-plane component, the dominant contribution to the Poynting flux will be in the two-dimensional horizontal plane. 
Shown in the right panels of Fig. \ref{fig:JandS} is $S_h$ a short time after the corresponding current density maps in the left panels. A striking feature of these maps of $S_h$ is the correspondence between the largest Poynting fluxes and strong concentrations of current density. Although not every intense current sheet appears as a region of strong flux, many do, and these correspond to locations of current dissipation and conversion of magnetic energy into plasma heating. Given these results, we use the locations of intense current sheets as proxies for locations where magnetic energy will be converted into heat, and large values of $S_h$ as a proxy for locations where energy dissipation is occurring. \par
\begin{figure*}
\centering\includegraphics[scale=0.25,trim=0.0cm 0.0cm 0cm 0.0cm, clip=true]{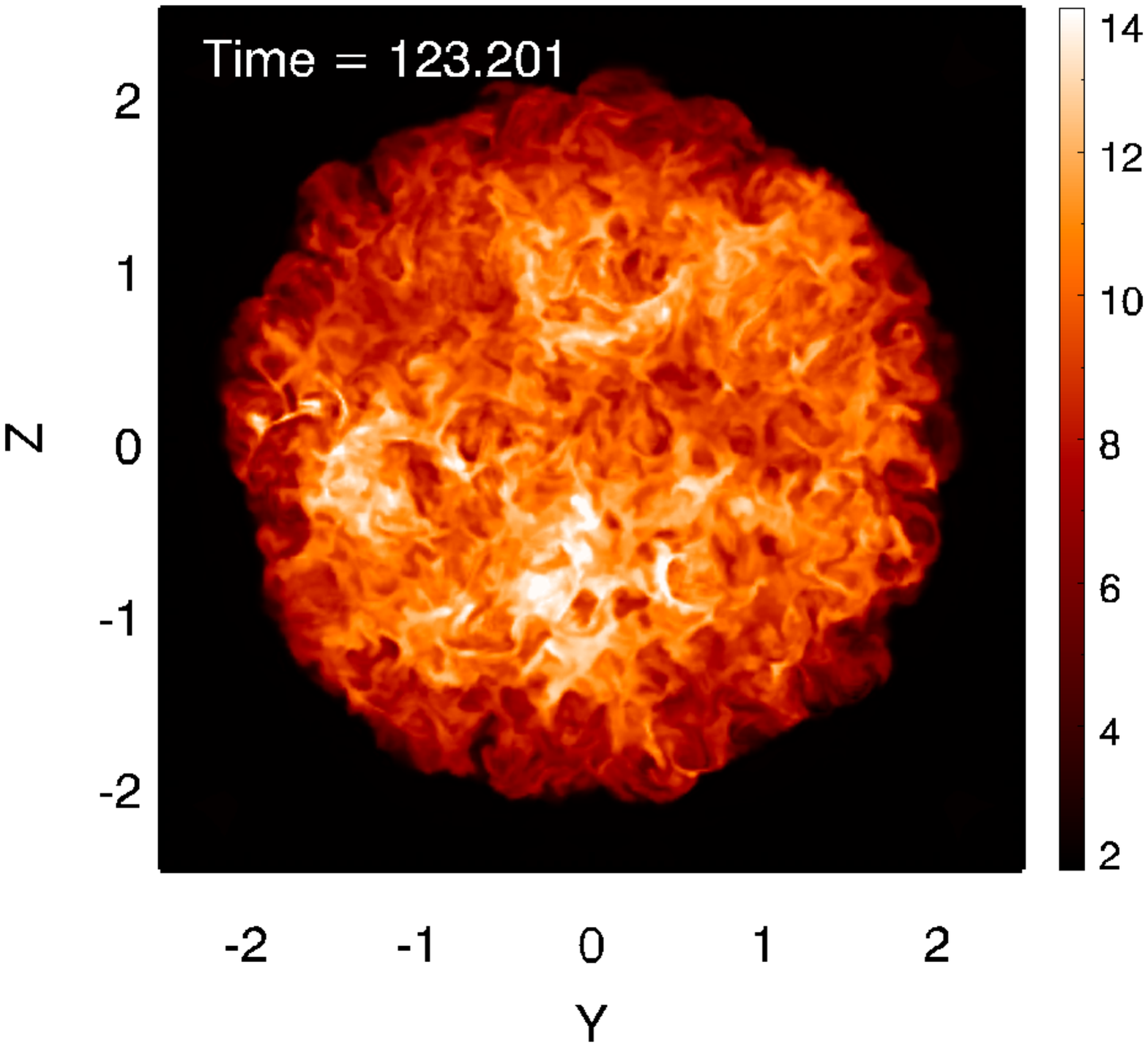}
\centering\includegraphics[scale=0.25,trim=0.0cm 0.0cm 0cm 0.0cm, clip=true]{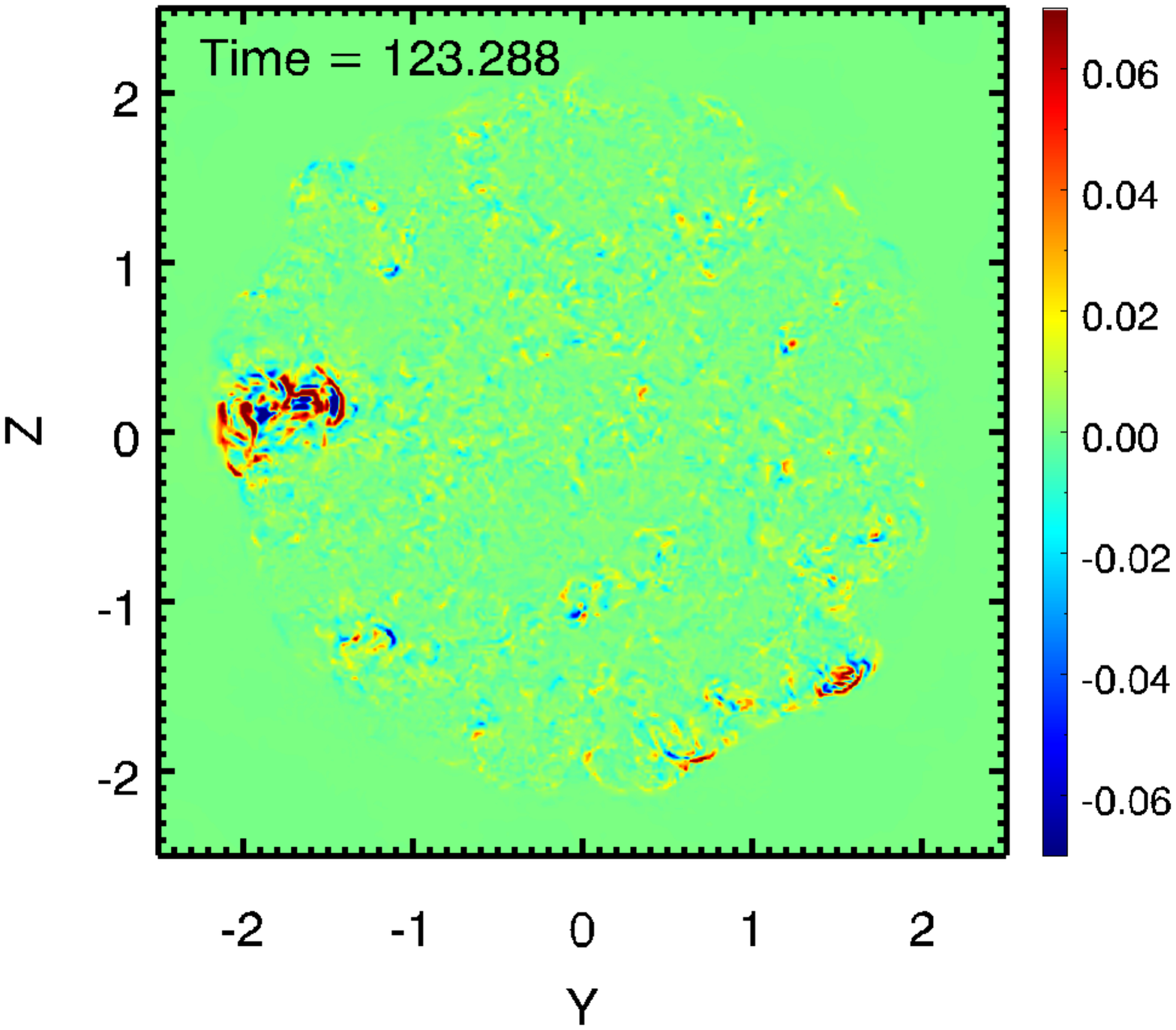}
\centering\includegraphics[scale=0.25,trim=0.0cm 0.0cm 0cm 0.0cm,clip=true]{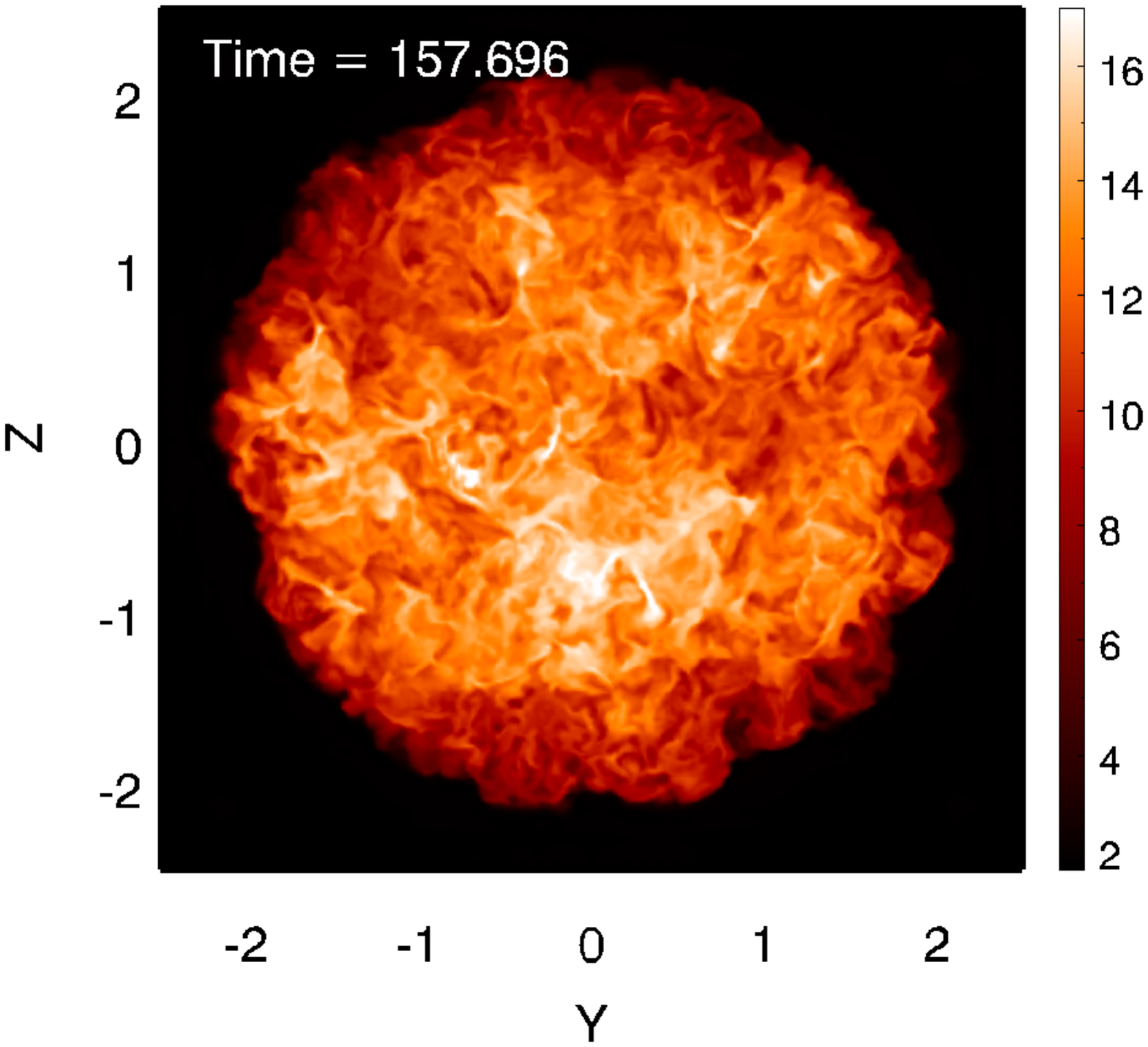}
\centering\includegraphics[scale=0.25,trim=0.0cm 0.0cm 0cm 0cm, clip=true]{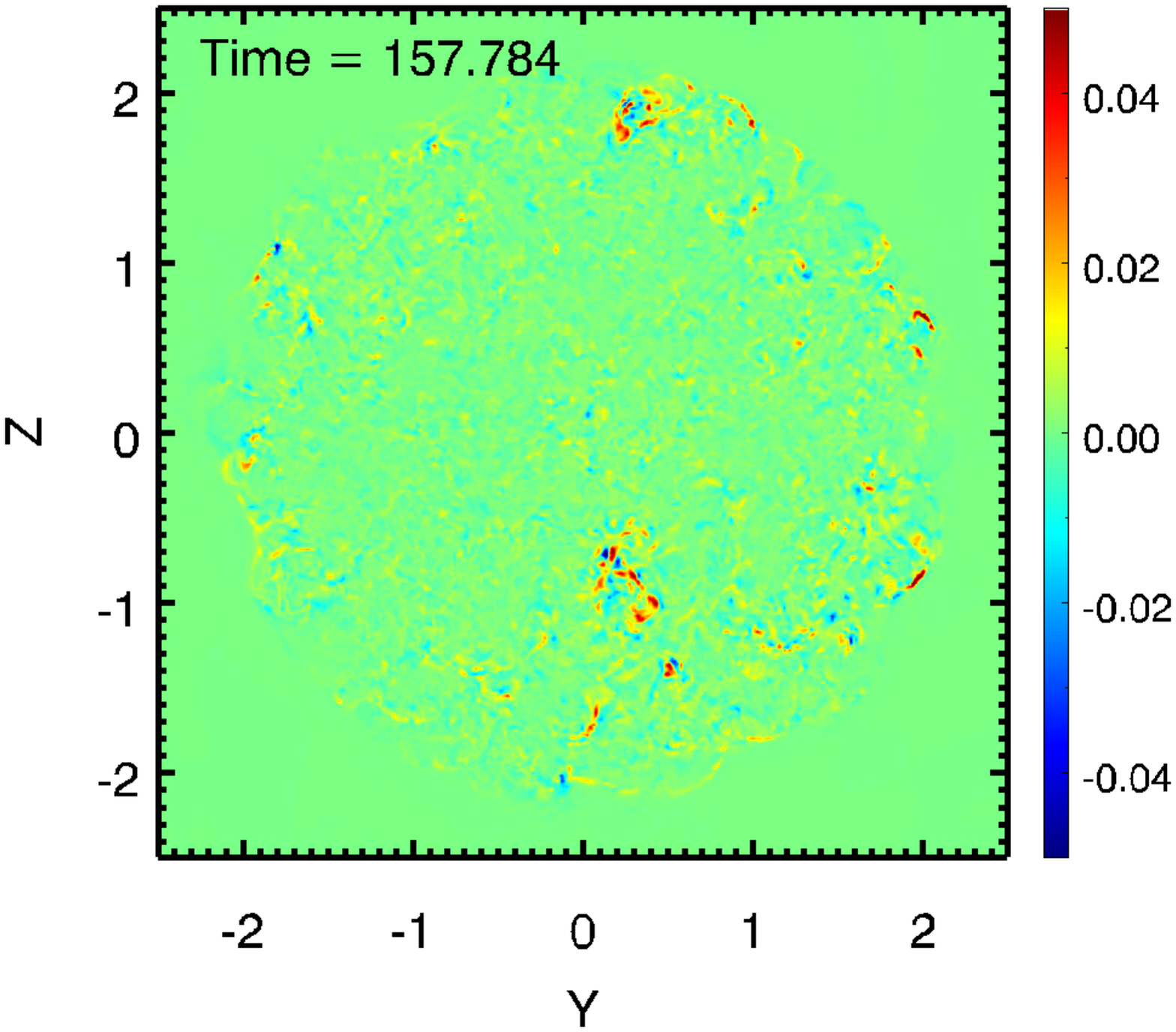}
\centering\includegraphics[scale=0.25,trim=0.0cm 0.0cm 0cm 0cm, clip=true]{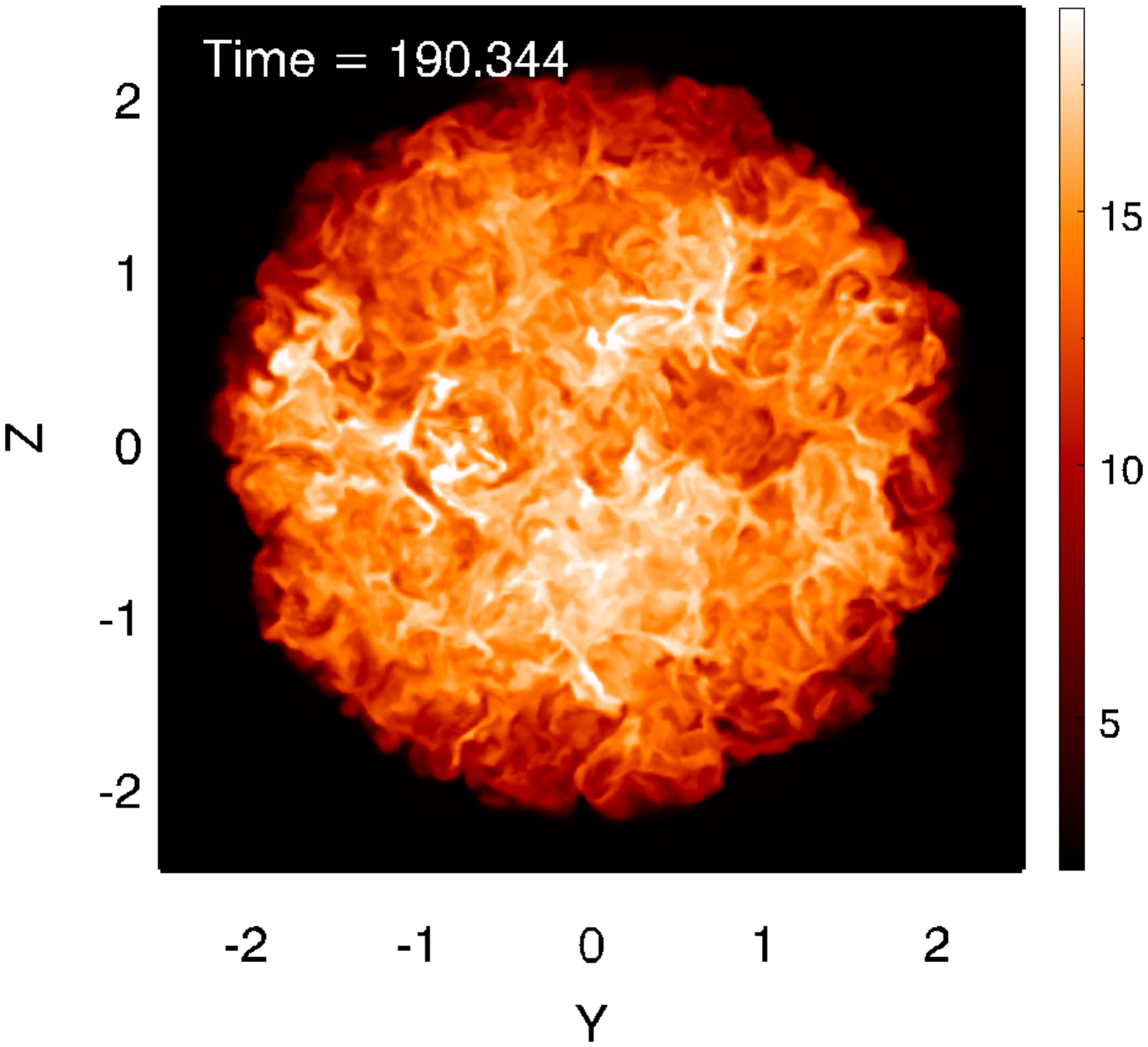}
\centering\includegraphics[scale=0.25,trim=0.0cm 0.0cm 0cm 0cm, clip=true]{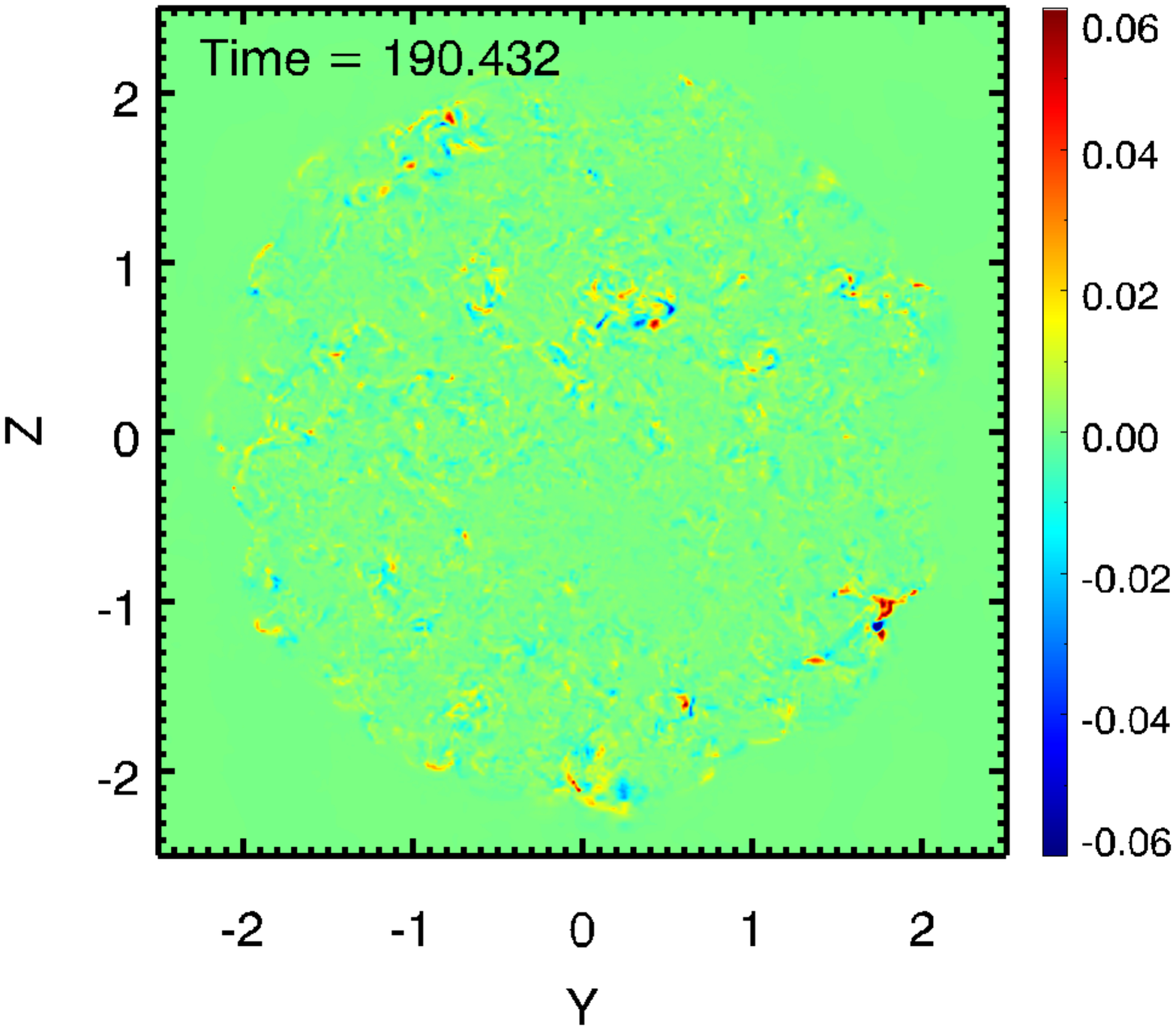}
\vspace{2mm}\caption{Left: Temperature in the midplane at three different times during the simulation, corresponding to the maps of $S_h$ above. The color scale is changing to account for the constant increase in temperature. Right: $\Delta T/T$ at the corresponding times.}
\label{fig:Temp}
\end{figure*}

The plasma heating immediately shows up in the temperature in the midplane. Fig. \ref{fig:Temp} shows the instantaneous temperature and change in temperature between two time steps, $\Delta T/T $. Since there is no thermal conduction or radiation in our model, plasma gets continuously heated without any significant cooling (though there are places where temperature decreases temporarily due to adiabatic expansion), so the temperature is constantly increasing, and locations where current sheets are dissipated may not be - at any given time - the hottest regions. Nevertheless, from the right panels in the figure it is obvious where the major heating events are occurring, as locations of largest change in temperature correspond very well spatially with locations of intense Poynting flux density. Note that examining the current density in the midplane would not necessarily have revealed these locations, since not every intense current sheet is undergoing reconnection at a given instant. The Poynting flux density, however, clearly shows locations where strong heating is occurring. 
\subsection{Statistics of Reconnection Events}\label{sec:statistics}
To analyze the statistics of our coronal model, we defined discrete reconnection events as contiguous regions in space and time characterized by an increased level of a relevant physical parameter distinguishing ``quiet'' and ``active'' states in the system. We selected three variables, taken over the course of the second half of the simulation, and processed them using a cluster identification code described below.\par


To characterize the bursty energy transport and dissipation associated with reconnection in our simulation, we used the difference of temperature $\Delta T (y,z,t) $ between subsequent time steps in the midplane of the model, the horizontal Poynting flux $S_h(y,z,t)$ throughout the midplane, and the midplane current density, $|\textbf{J}(y,z,t)|$. Events in all three data fields were identified using the algorithm by \citet{Uritsky10} and \citet{Uritsky10JGR}. The first step of the algorithm consists of identifying times and locations of the grid nodes in which the studied parameter exceeds a predefined activity threshold, expressed as a given number of standard deviations above the mean. The initial and final times as well as the locations of each activity burst above the threshold are recorded. Second, the activity bursts are labeled according to their spatial and temporal connectivity, with events occurring in neighboring grid nodes and overlapping in time having the same label.
For this purpose, we considered sets of 8 nearest grid nodes on the 2D midplane, including the diagonal neighbors. The labeling process uses the ``breadth-first search'' principle to avoid backtracking of search trees representing individual clusters. Finally, the activity bursts are sorted according to their labels, and the spatiotemporal domains of each event are obtained. Events occurring at the perimeter of the midplane or that are truncated by the finite time frame of the analysis are removed from the statistics. See \citet{Uritsky10} for more details on the method.

The following statistical characteristics have been computed for each detected event: its area, defined as the largest number of grid nodes in the midplane simultaneously involved in the event,
\beg{area}
\mathcal{A}_i = max(A_i) = max( \delta^2 \sum_{k\in\Lambda_i}{k})\; ;
\done
its lifetime, given by the difference of its ending and starting times,
\beg{lifetime}
\tau_i = t_f-t_i \; ;
\done
its spatiotemporal volume, given by a combination of area and lifetime,
\beg{STV}
V_i = \sum_{t_i}^{t_f}{A_i}\; ;
\done
and its ``energy,'' obtained by integrating the studied parameter ($\Delta T$, $S_h$, or $|\textbf{J}|$) over the entire spatiotemporal domain of the event, considering its exact 3D shape,
\beg{energyevent}
E_i = \sum_{k\in\Lambda_i}{C_k}\;.
\done
Here $\delta=\sqrt{dy^2+dz^2}$ is the uniform grid spacing, $k$ labels the grid point, $\Lambda_i$ is the set of all grid points involved in the $i^{th}$ event, $t_f$ and $t_i$ are the final and initial times of the event, respectively, and $C_k$ is the parameter being evaluated at the $k^{th}$ node. Each event can, in general, start and end at different grid points. For each of the parameters, we approximate the distribution via a power law form:
\beg{powerlaw}
p(X) \sim X^\alpha
\done
where $p(X)$ is the occurrence rate of characteristic $X=\mathcal{A}_i$, $\tau_i$, $V_i$, or $E_i$ \citep{Uritsky10}.

\begin{figure*}
\centering\includegraphics[scale=1.0, trim=3.0cm 17.0cm 10.0cm 3.0cm,clip=true]{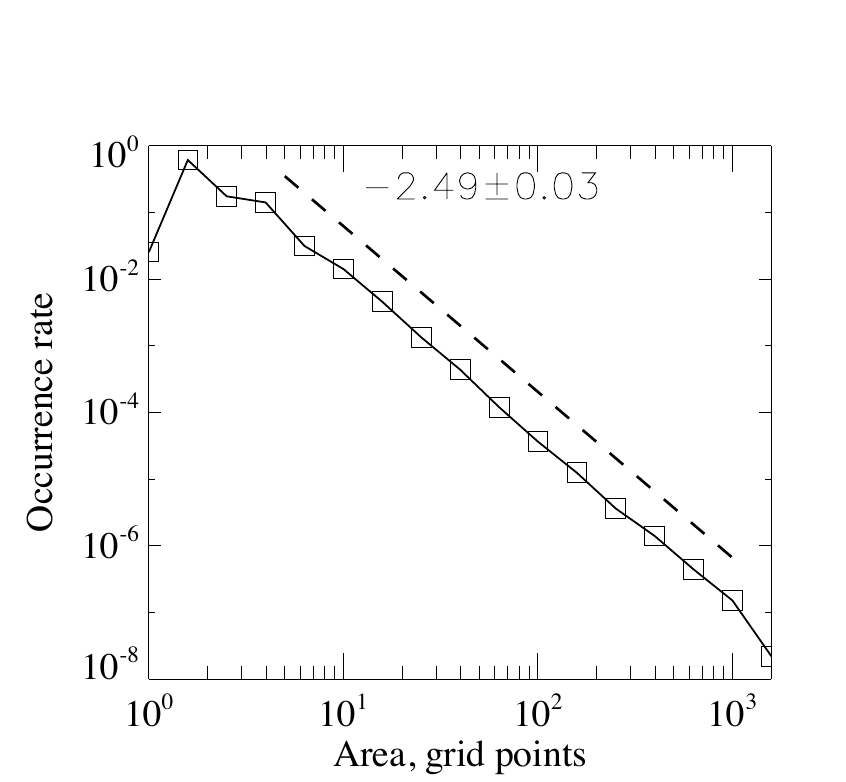}
\centering\includegraphics[scale=1.0, trim=3.0cm 17.0cm 10.0cm 3.0cm,clip=true]{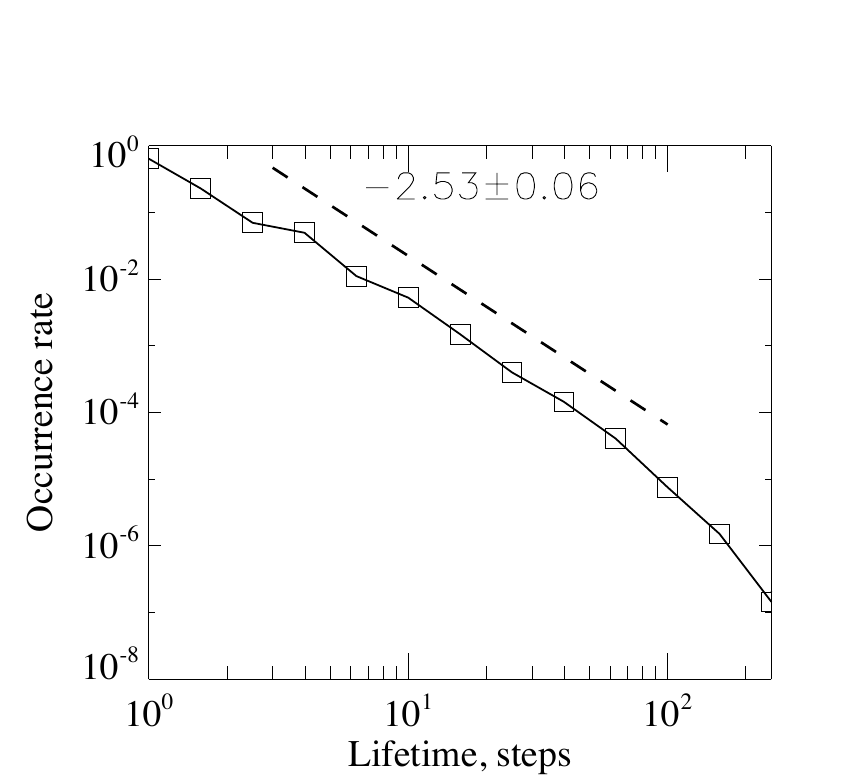}
\centering\includegraphics[scale=1.0, trim=3.0cm 17.0cm 10.0cm 4.0cm,clip=true]{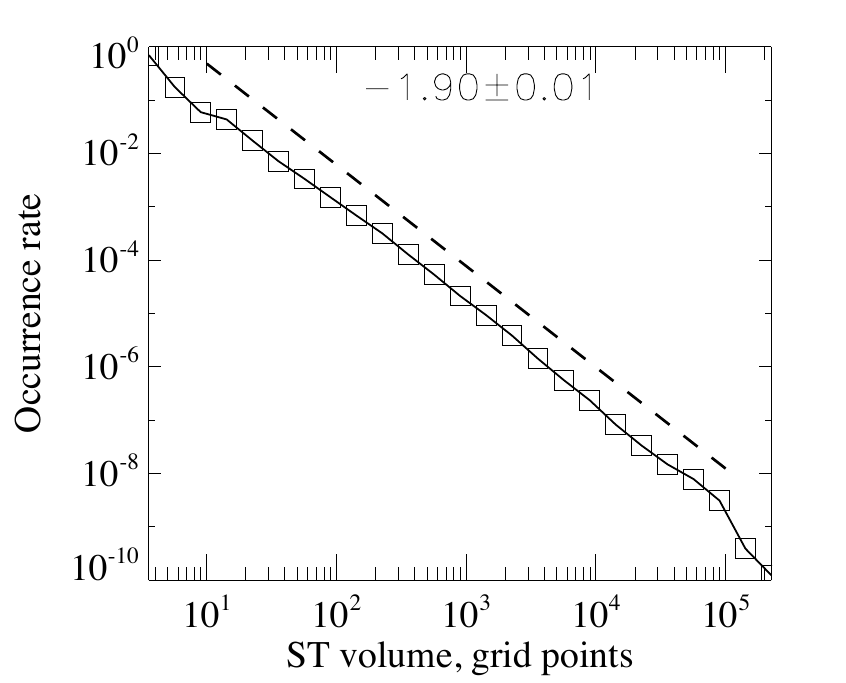}
\centering\includegraphics[scale=1.0, trim=3.0cm 17.0cm 10.0cm 4.0cm,clip=true]{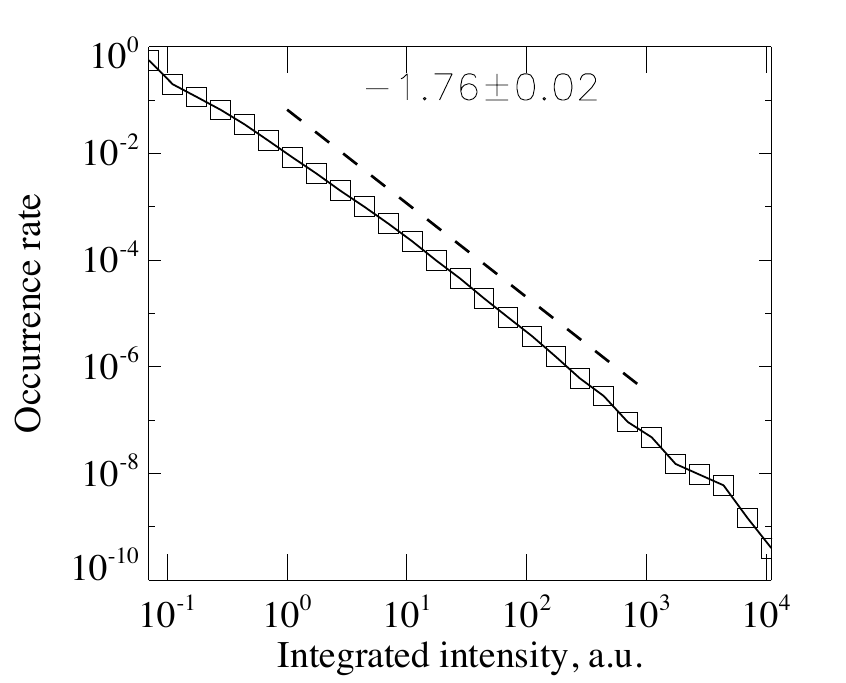}
\caption{Probability distribution of temperature difference $\Delta T$ as a function of area (top left), time (top right), area+time (bottom left), and energy (bottom right).}
\label{fig:psdT}
\end{figure*}

Figure \ref{fig:psdT} shows the relative occurrence rate of strictly positive incremental changes in temperature $\Delta T$ more than two standard deviations above the mean. The occurrence rates are calculated as a function of the four event characteristics listed above. The distributions of temperature increments exhibit broad-band power laws with well-defined slopes. The mean values and the standard errors of the power-law exponents provided on each panel are obtained using linear regression analysis in log-log coordinates. The range of scales of the event activity in this parameter (Figure \ref{fig:psdT}) can be read off the two-dimensional area-occurrence rate plot, which shows a power-law slope of $-2.49$. Taking the square root of the lower and upper bounds of the linear portion of the log-log fit gives the spatial scale, in grid points, of the largest and smallest scales. In particular, the upper bound corresponds to a length scale of $\sqrt{10^3} \approx 32$ grid points, which indicates that the largest heating events occur on the scale of the convection cells, while the lower bound indicates a scale of just a couple grid points, which indicates that the dissipation scale is essentially the grid spacing. Although the occurrence rate of temperature changes with lifetime follows a power law over only a decade, the spatio-temporal volume and absolute magnitude curves show a clear power law behavior over more than three decades.\par

\begin{figure*}
\centering\includegraphics[scale=1.0, trim=3.0cm 17.0cm 10.0cm 3.0cm,clip=true]{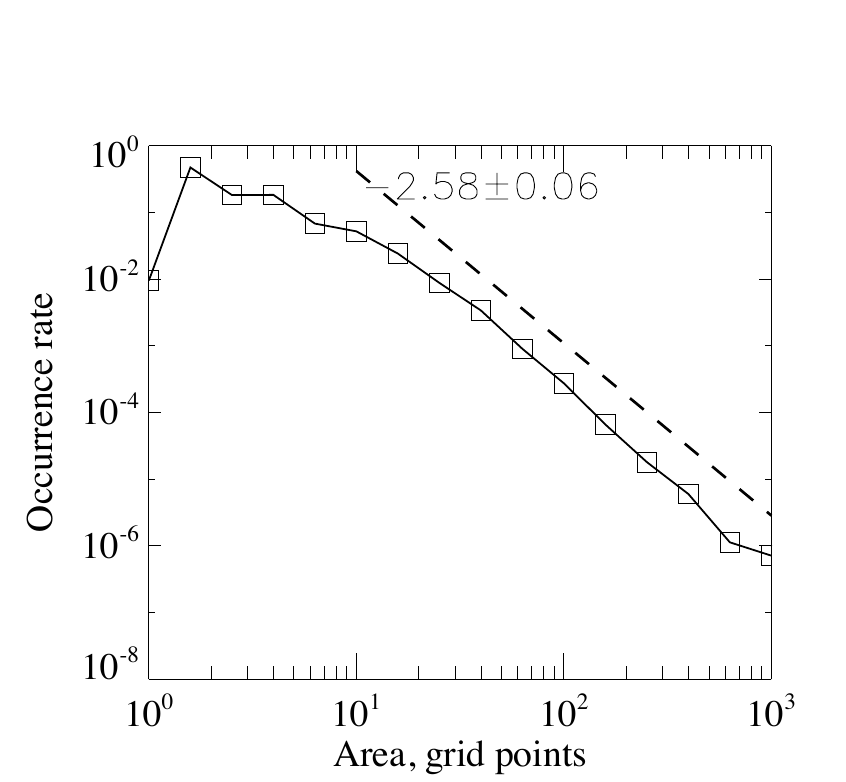}
\centering\includegraphics[scale=1.0, trim=3.0cm 17.0cm 10.0cm 3.0cm,clip=true]{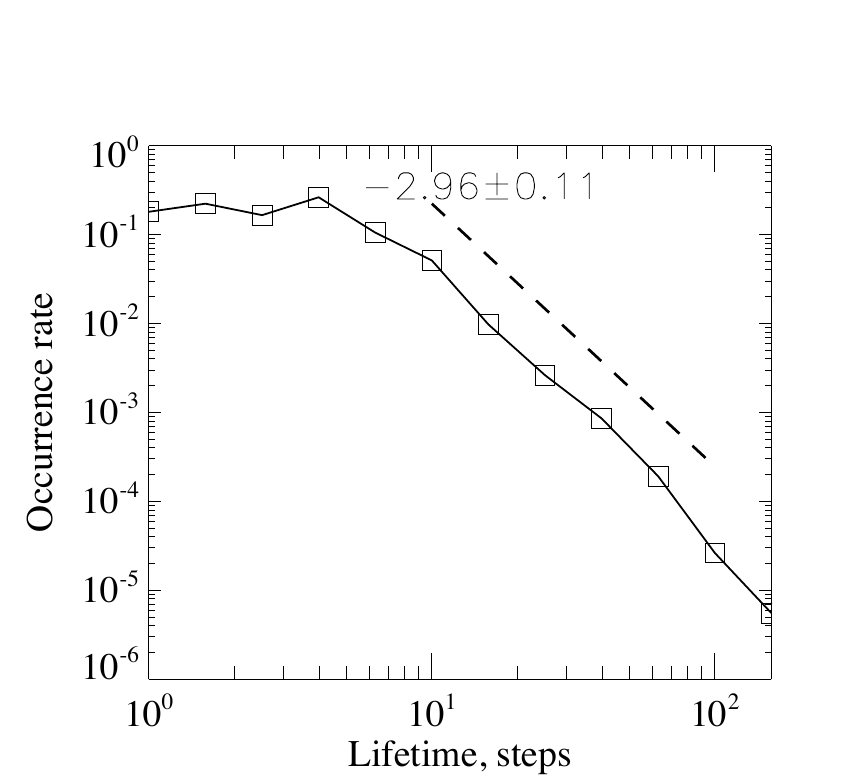}
\centering\includegraphics[scale=1.0, trim=3.0cm 17.0cm 10.0cm 4.0cm,clip=true]{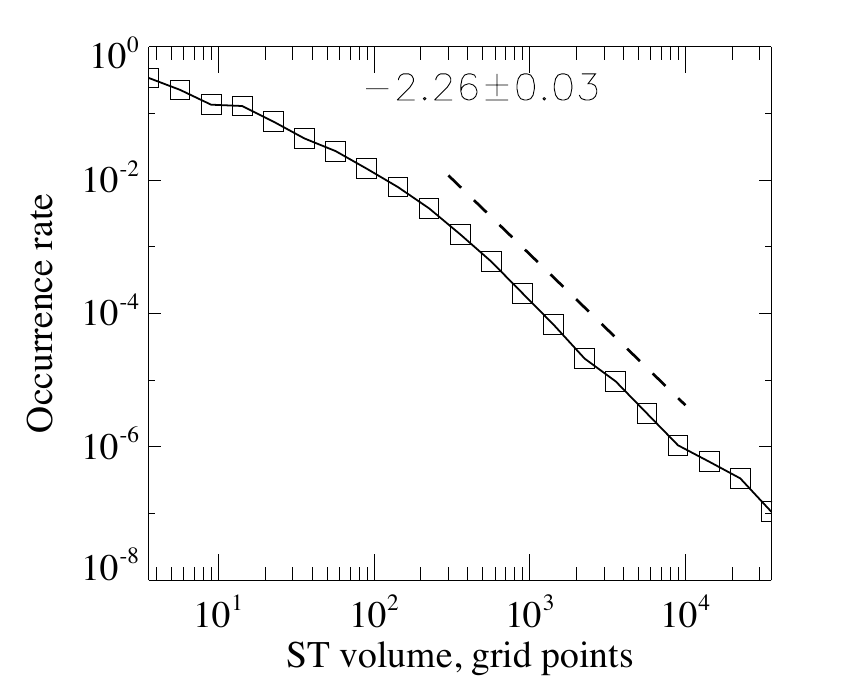}
\centering\includegraphics[scale=1.0, trim=3.0cm 17.0cm 10.0cm 4.0cm,clip=true]{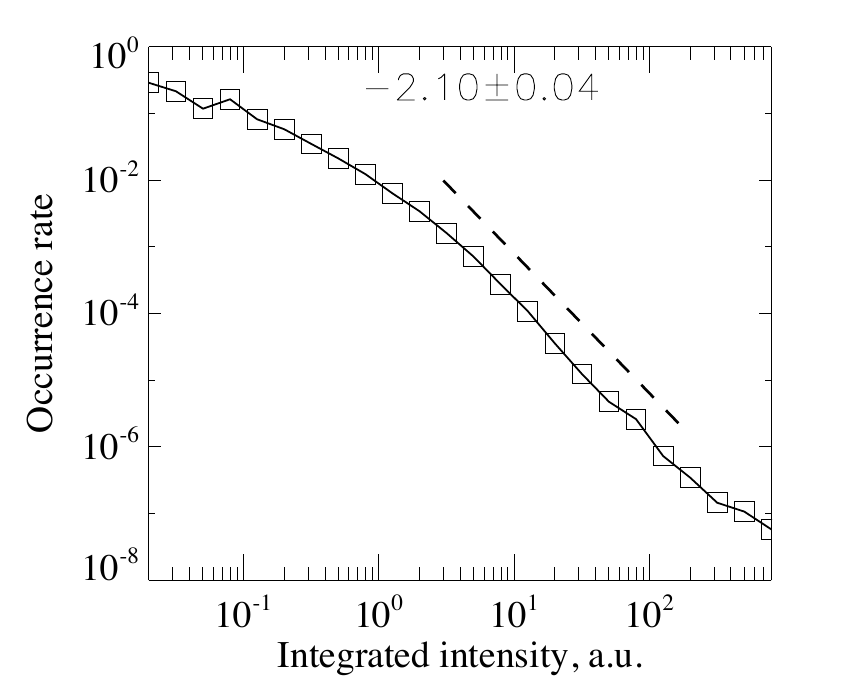}
\caption{Probability distribution of horizontal Poynting flux as a function of area (top left), time (top right), area+time (bottom left), and energy (bottom right).}
\label{fig:psdS}
\end{figure*}

Figure \ref{fig:psdS} shows the relative occurrence rate of horizontal Poynting flux fluctuations more than two standard deviations above the mean.  While the Poynting flux is much more noisy and intermittent than temperature changes, there is, nevertheless, a clear indication of a power law behavior over a range of scales of at least one order of magnitude.\par

\begin{figure*}
\centering\includegraphics[scale=1.0, trim=3.0cm 17.0cm 10.0cm 3.0cm,clip=true]{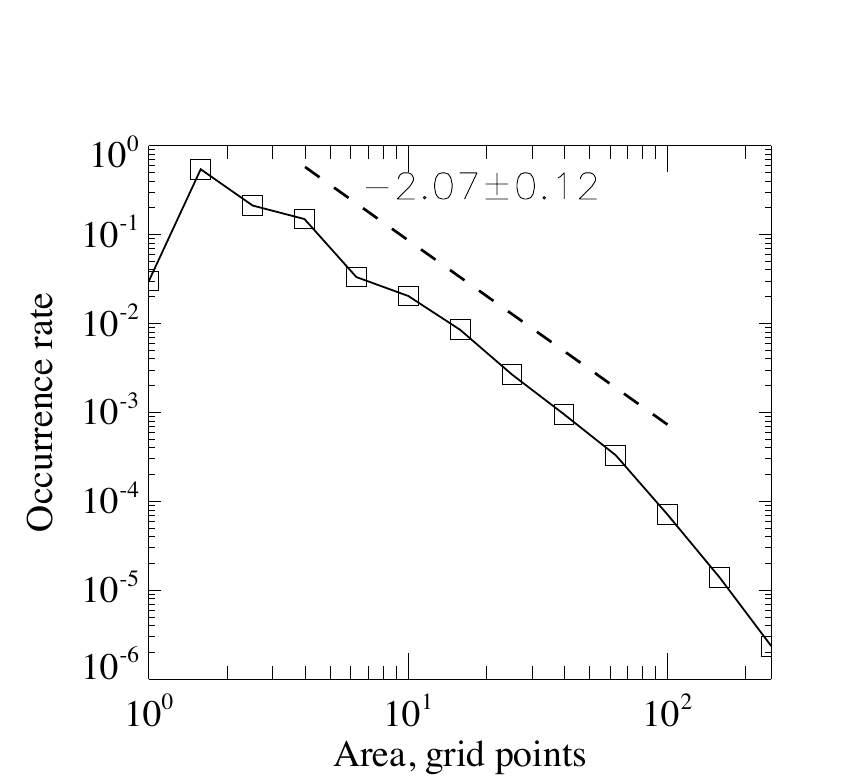}
\centering\includegraphics[scale=1.0, trim=3.0cm 17.0cm 10.0cm 3.0cm,clip=true]{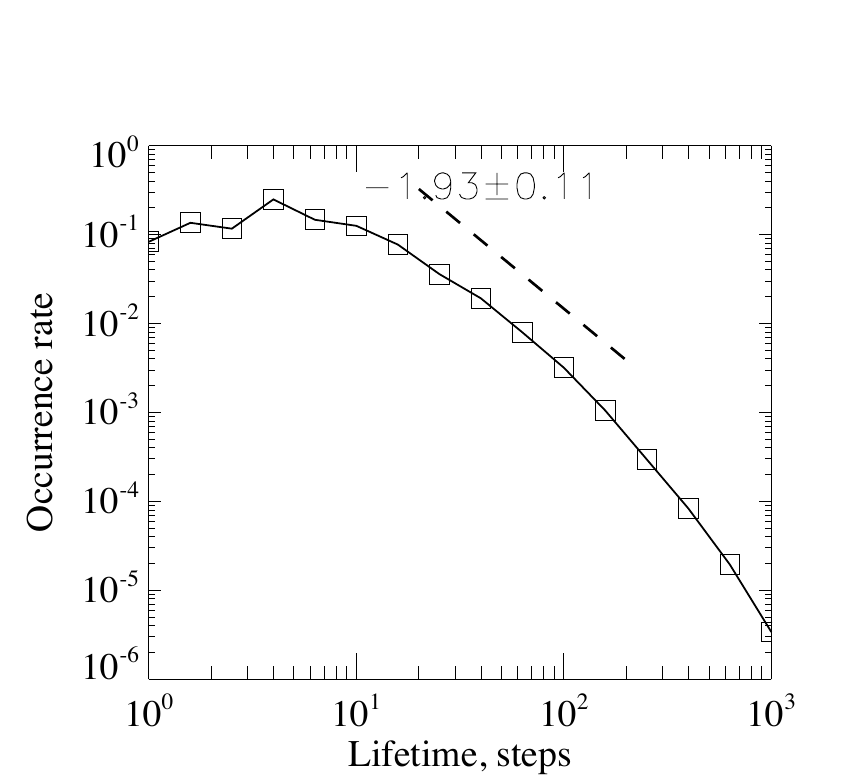}
\centering\includegraphics[scale=1.0, trim=3.0cm 17.0cm 10.0cm 4.0cm,clip=true]{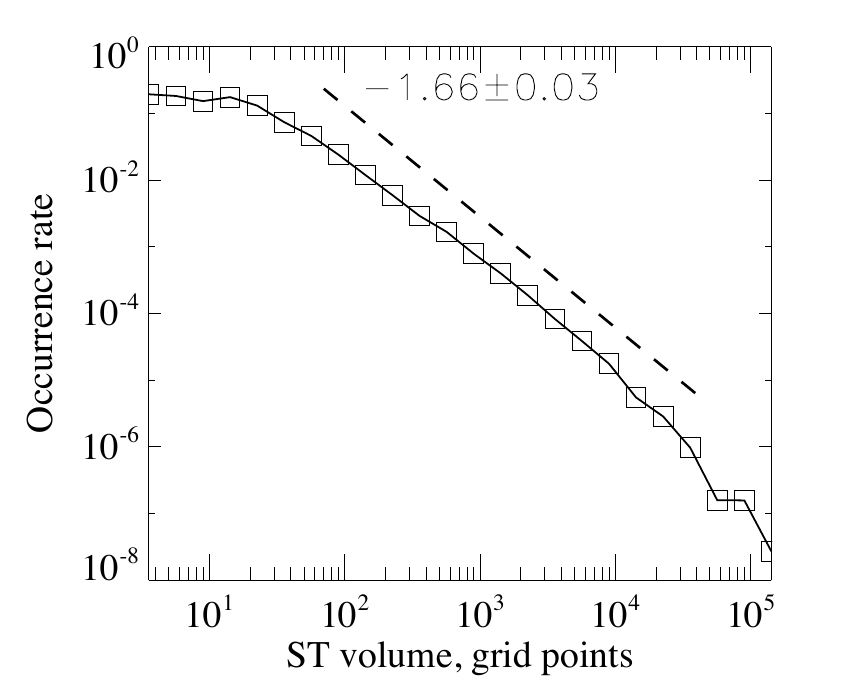}
\centering\includegraphics[scale=1.0, trim=3.0cm 17.0cm 10.0cm 4.0cm,clip=true]{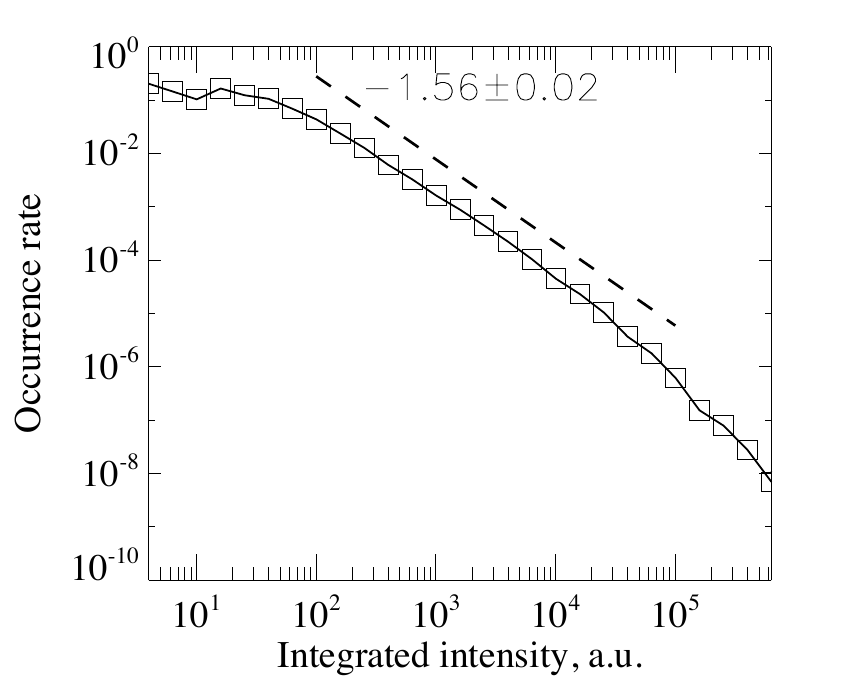}
\caption{Probability distribution of current density as a function of area (top left), time (top right), area+time (bottom left), and energy (bottom right).}
\label{fig:psdJ}
\end{figure*}

Figure \ref{fig:psdJ} shows the relative occurrence rate of current density fluctuations more than two standard deviations above the mean. There is a clear power-law distribution evident over several decades in scales, with slopes ranging from $\approx -1.5$ to $\approx -2.1$. \par 

The power-law slopes obtained for the parameters studied here are qualitatively and quantitatively similar not only to those of toy-model cellular-automaton systems such as those of \citet{Lu91}, but also with observed values. Denoting with $\alpha_P$, $\alpha_E$, and $\alpha_t$ the absolute value of the observed slopes of the power laws for peak flux, energy, and duration, respectively, these distributions have been measured to have $1.58< \alpha_P <2.00$, $1.39<\alpha_E<1.74$, and $1.88<\alpha_t<2.4$, respectively, in the hard X-rays. Solar flare soft X-ray emission display power-law distributions with slopes $1.64<\alpha_P<2.16$ $1.44<\alpha_E<2.03$, and $2.02<\alpha_t<2.93$. In the EUV, $1.19<\alpha_P<2.66$, $1.41<\alpha_E<2.31$, and $1.4<\alpha_t<2.74$ \citep{Aschwanden16}. These are all in approximate agreement with the results obtained here.

\subsection{Magnetic Field Spectrum}
\begin{figure*}
\centering\includegraphics[scale=0.5, trim=0.0cm 0.0cm 0.0cm 0.0cm,clip=true]{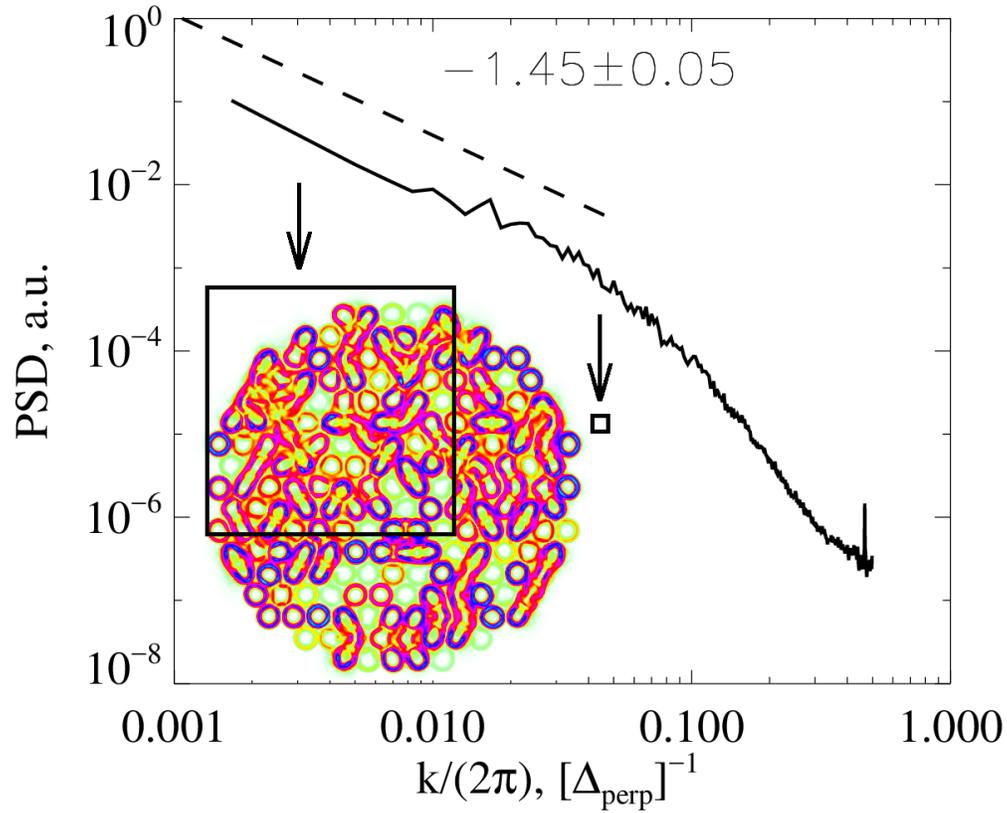}
\caption{Probability distribution of magnetic field for the last half of the simulation, along with a rendering of the 2D cut of the transverse magnetic field at the end of the simulation. The big box shows the approximate scale of $\Delta_{perp}^{-1} = 0.003$, while the little box shows the approximate scale of $\Delta_{perp}^{-1} = 0.04$, where there is a break in the spectrum.}
\label{fig:psdB}
\end{figure*}

An alternative way of quantifying intermittent energy dissipation in plasmas involves a power spectral analysis approach. Low-beta plasmas dominated by magnetic pressure forces can be characterized by the Fourier power spectrum of the horizontal magnetic field.  Figure \ref{fig:psdB} shows the Fourier spectrum of magnetic field magnitude in the midplane. The vertical axis shows the spectral density of the horizontal magnetic field with the spatial scale given by $\Delta_{perp}^{-1}$. The inset shows a snapshot of the transverse magnetic field magnitude at one of the time steps included in the spectral analysis. 

\begin{figure*}
\centering\includegraphics[scale=1.0, trim=3.0cm 17.0cm 10.0cm 3.0cm,clip=true]{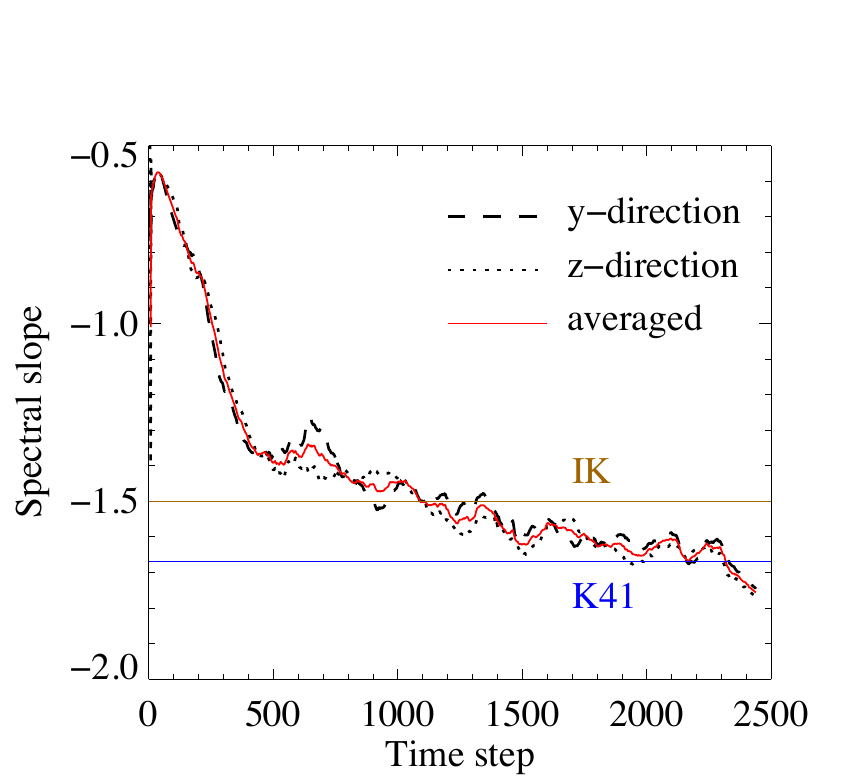}
\centering\includegraphics[scale=1.0, trim=3.0cm 17.0cm 10.0cm 3.0cm,clip=true]{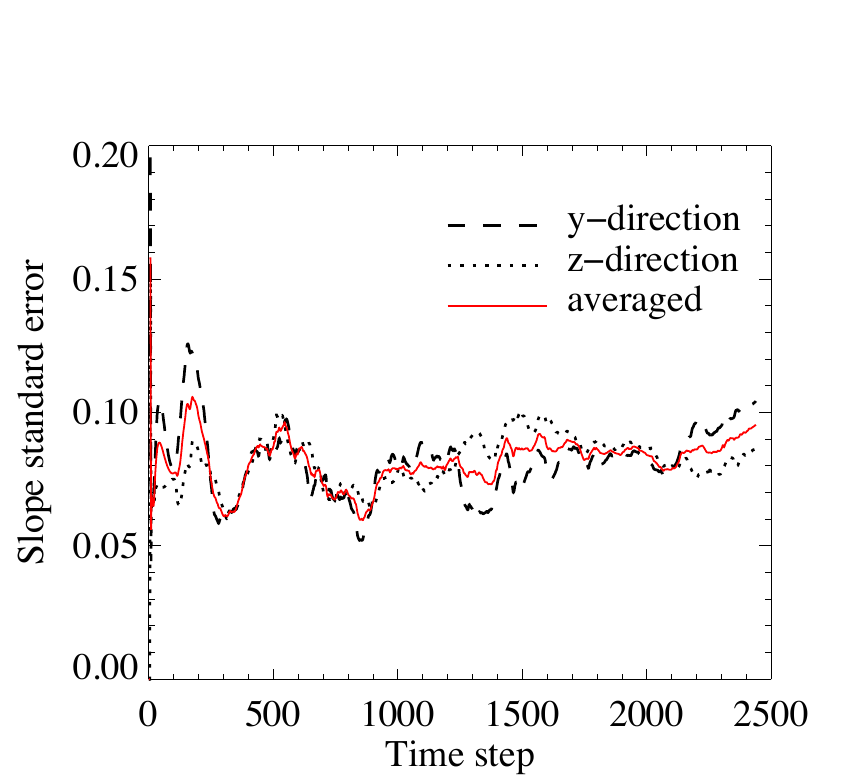}
\caption{Spectral slope (left) and its uncertainty (right) as a function of time. The spectral slope has been calculated by integrating over both transverse directions ($y,z$) and taking their average (red curve). Overplotted for reference are the IK (gold) and K41 (blue) slopes.}
\label{fig:turbulence}
\end{figure*}

The obtained magnetic spectrum contains two intervals of power law behavior described by distinct log-log slopes. The large-scale interval occupying the transverse linear scales ranging from a single convective cell (in agreement with the inference from Figure \ref{fig:psdT}) to about half the system size (shown with small and large squares, respectively) has a slope of about $-1.45$. Analysis of the time-dependent behavior of the power law slope of the horizontal magnetic field, shown in Figure \ref{fig:turbulence}, provided additional insights into the development of the state of the model. The power-law exponent of the magnetic spectrum seems to behave in a fairly nonstationary way over the course of the simulation, oscillating near $-1.45$, but approaching $\approx -1.67$ towards the end of the simulation. The standard error of the the log-log slope shown on the right panel of the figure can be seen as a measure of the quality of the power-law model, which varies within broad limits. Similar results were obtained when computing spectra from the signed horizontal components of the magnetic field. 

The power-law slope seen in Figure \ref{fig:psdB} agrees approximately with the \citet{Iroshnikov63} and \citet{Kraichnan65} (IK) prediction for an MHD turbulent cascade over a decade and a half in wavenumber. In the IK framework, the interaction of counter-streaming Alfv\'en wave packets traveling along the magnetic field is responsible for dissipating energy in a scale-free manner. On the other hand, we observe an apparent break in the spectrum at $\Delta_{perp}^{-1} = 0.04$ above which the spectrum decays much faster. The position of the spectral break corresponds to about $25$ grid cells in scale, in close agreement with the number of grid cells across the diameter of each one of our convective cells. Although the measured spectral slopes are roughly consistent with the \citet[][K41]{Kolmogorov41} or IK interpretations, it is also possible that the generation of the large-scale horizontal magnetic field in our model is qualitatively different from either turbulence scenario. Because of the presence of localized reconnection regions, the smallest scales in the  simulation can operate not only as energy sinks, but also as energy sources by launching inverse cascades toward larger scales \citep{Uritsky09}. This type of stochastic behavior lies beyond the scope of traditional turbulence theories, and it can play a significant part in many space plasma systems with complex reconnecting magnetic topologies. Consequently, the results presented here provide evidence for multiscale, stochastic magnetic structures exhibiting two distinct scaling regimes, with a crossover at the transverse length scale corresponding to a single convection cell. However, it cannot be stated definitively what type of turbulence is exhibited by this system, or, indeed, whether it is truly turbulent at all. The reconnecting current sheets and the spaces between convective cells are spanned by only a few grid points, very close to the smallest scales in the system where the energy is ultimately dissipated. \par 


\section{Discussion}\label{sec:implications}
In this work, we presented a model for the continuous photospheric driving of the coronal magnetic field that is designed to mimic the salient observational features of convection, including its randomness, helicity injection preference, spatial compactness, and temporal profile. We used several proxies for energy release -- the change in temperature, the horizontal Poynting flux, and current density -- to study the response of the coronal field to this photospheric driving. We found that energetic events in the model coronal field exhibited power-law distributions over several decades in scales.  Further, we analyzed the magnetic spectrum occurring in our simulation and found that, for spatial scales larger than a convection cell, the power-law slopes are consistent with IK turbulence for the majority of the simulation and Kolmogorov turbulence towards the end, but that the details of our model prevent a positive identification of the evolution of our system as one or the other. The presented statistics show that energy dissipation in our model occurs in a form of sporadic spatiotemporal activity bursts which are qualitatively similar to the statistics of solar flares obtained from observations \citep[e.g.][and references therein]{Crosby93, Benz02, Aschwanden12,Uritsky07,Uritsky13,Uritsky14,Aschwanden16}. This behavior unfolds over transverse spatial scales of the order of several adjacent flux tubes and represents reconnection events on the misaligned field lines of these tubes. Our model shows results consistent with other models coupling the photospheric motions to coronal flaring statistics \citep[e.g.,][]{Nigro04,  Uritsky13, Mendoza14}, but the key difference with these works is the drastically different scenario in which a complex turbulent flow is not required for multiscale flaring activity. Instead, the boundary flow applied here is quite simple, yet the resulting flaring is quite complex and multiscale, in agreement with coronal observations of power-law probability distributions. Furthermore, the power-law slopes we obtain are well in the range of the observed power-law slopes observed in hard and soft X-ray spectra as well as EUV emission. \par
Our work is the first investigation of 
spatio-temporal
power-law statistics in a fully $3$D MHD, driven coronal-loop simulation. 
Our results are both qualitatively and quantitatively consistent with the results of \citet{Kanella17}, who used the \emph{Bifrost} code with radiative and conductive processes to study heating statistics in a model solar atmosphere driven by convective flows from below their bottom boundary. They analyze the statistics of Joule heating events, a direct output from their code, using a feature-identifying algorithm to determine the power-law spectra of energy release sizes and rates. For a single timestep during their simulation, they identify several thousand events, whose distributions follow a power law with spectral indices in the $1.5-2.5$ range, in good quantitative agreement with the results from the heating proxies presented here. A key difference of our study is the finding that reconnection events follow a temporal, in addition to spatial, power-law distribution. This finding has major implications for the solar corona, since the temporal robustness of the statistical result implies that the corona should constantly be releasing energy in a well defined power-law distribution, and not just at a fortuitously selected instant.
\par
The results of our simulations suggest that power-law statistics should be an ever-present feature of the solar corona. Such power-laws have been demonstrated in toy models such as \citet{Lu91}, and the results presented here are fully consistent with those models and put them on much firmer physical footing. Driven MHD current sheets have been shown to exhibit power-law statistics in the context of the Earth's magnetosphere \citep{Klimas04}. A critical question, however, is: Is the scale-free behavior seen in our model indicative of instabilities that generate disturbances far away from the initial energetic event? Many simulations have been performed in which enough stress has been injected at the photospheric level to trigger the onset of ideal instabilities, such as the kink mode in individual flux tubes \citep[e.g.,][]{Torok03,Zhao15}. These instabilities could allow a localized event to trigger successive energy releases far away from the location of the initial instability. \citet{Hood16} proposed such a reconnection induced `avalanche' model in which a localized instability triggers a large scale energy release. Starting with a single kink-unstable flux tube surrounded by a pattern of marginally stable twisted flux tubes, they initiated an MHD avalanche. Successive reconnection events caused the initial kink-unstable flux tube to interact with, and influence, the majority of the other flux tubes. This was the first reported simulation of an MHD avalanche. \par
For such an avalanche to occur, there must be significant free energy in the coronal field, since otherwise a small number of reconnection events between neighboring tubes will be sufficient to bring the system back to its lowest energy state. On the other hand, injecting significant twist by driving an initially untwisted field will quickly cause current sheets to develop between adjacent flux tubes, whence reconnection will transport that twist far away from the injection scale \citep{Antiochos13, Zhao15, Knizhnik15, Knizhnik17}. Thus, rather than a localized event causing disturbances far away, many localized events combine to influence the system on a global scale. As a result, these events will not be causally related to each other. \par
The statistical behavior seen in our model is suggestive of avalanche behavior typically observed in self-organized critical (SOC) states \citep{Bak87} previously seen in driven \citep{Dmitruk98, Klimas04} and undriven \citep{Hood16} MHD systems. In an SOC state, the system reaches a critical point of a phase transition, but, crucially, without external tuning of parameters. The sandpile model is a famous example, where sand is continuously dropped onto a pile, until the pile reaches a critical state, at which point the addition of any more sand will cause avalanches that readjust the shape of the sandpile. The size distributions of such avalanches typically obey power laws, although these have really only been achieved self-consistently in cellular automaton models \citep{Bak88, LF15,LF16}, or $2$D MHD simulations \citep{Dmitruk98, Georgoulis98} \citep[although see, for example, ][and references therein]{Klimas17}. While these cellular automaton models, and the power laws they exhibit, suggest that the coronal magnetic field approximates a SOC state \citep{Lu91,Vassiliadis98,Charbonneau01,Krasnoselskikh02,Dimitropolou13,Uritsky13}, statistical behavior does not itself prove the presence of SOC, but power-law distributions are commonly considered a necessary condition for SOC \citep{Uritsky09}. \par
\begin{figure*}
\centering\includegraphics[scale=0.65, trim=0.0cm 7.0cm 0.0cm 0.0cm]{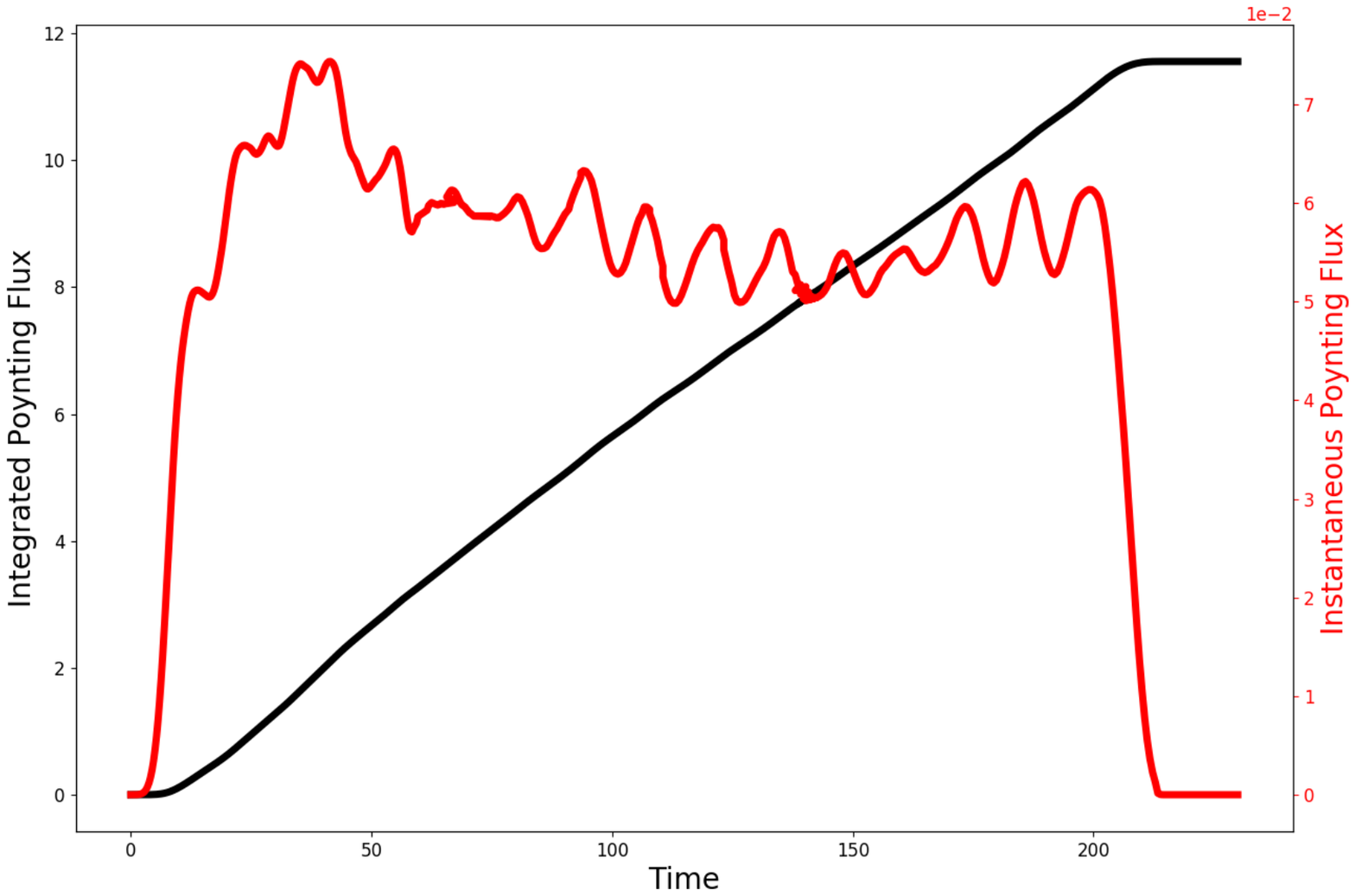}
\centering\includegraphics[scale=0.65, trim=0.0cm 5.0cm 0.0cm 7.0cm]{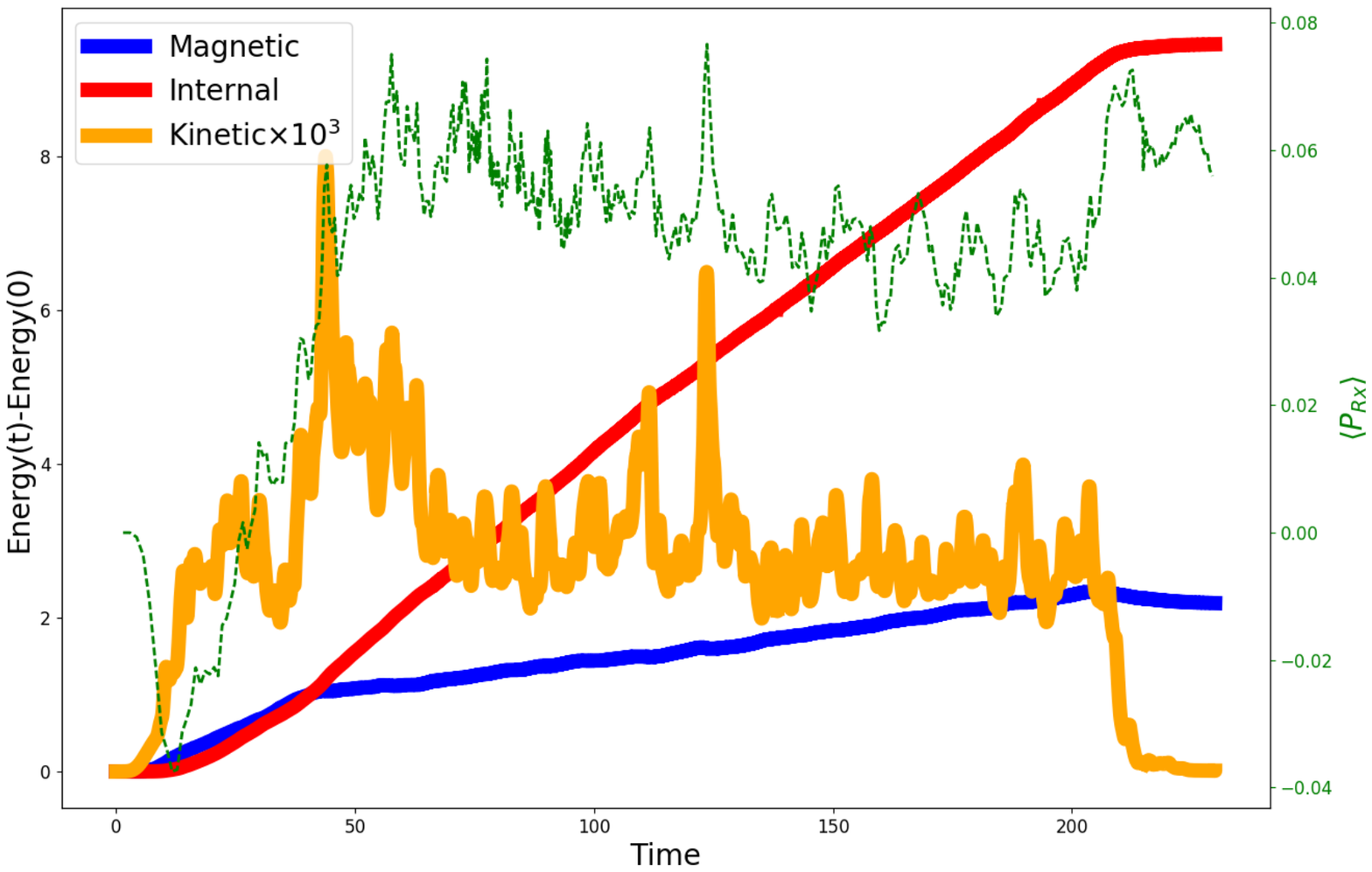}
\caption{\textbf{Top:} Instantaneous Poynting flux (red) and time integrated Poynting flux (black). 
Bottom: Evolution of magnetic (blue), plasma (red), and kinetic ($\times 10^3$; orange) energies as a function of time, as well as the energy conversation rate (dashed green).
}
\label{fig:energies}
\end{figure*}

 
Typical SOC models reach a steady state in which energy injection is, on average, balanced by energy dissipation. To determine whether our model is truly in an SOC state requires a full treatment of conductive and radiative effects. Although our model conserves energy, it does not include these important terms in the energy equation. As a result, our model lacks some of the key physics to definitively address this point. Nevertheless, \citet{Vespignani98} suggest that an important property of SOC systems is that the energy conversion rate of localized reconnection events needs to reach a steady state. The average volume energy conversion rate, shown in Fig. \ref{fig:energies}, has approximately constant periodicity and oscillation amplitude. This provides indirect evidence that, even though the global steady-state has not been reached, the reconnection rate and the associated avalanching activity can be considered quasi-stationary, as required for an SOC state. 
\par
The nature of the photospheric energy injection makes modeling MHD avalanches in the corona very challenging. The fact that the energy, in the form of twist, is injected in narrow lanes means that the azimuthal magnetic field generated as a result of this motion will act as a destabilizing force against the stabilizing axial magnetic field, leading to the rapid development of kink modes \citep[e.g.][]{Anzer68}, which triggered the MHD avalanche seen in \citet{Hood16}. On the other hand, the proximity of the photospheric injection cells to each other may be expected to cause reconnection between adjacent flux tubes to occur before the kink threshhold is reached, which could lead to a much less explosive energy release. Thus, the photospheric driving of the coronal magnetic field is conducive to continuous bursts of energy release, but may not necessarily be able to cause other energetic events far away from the original site. Future work will investigate whether such `avalanches' can truly occur in driven systems.\par

\acknowledgments{
We are grateful for the comments of the anonymous referee, whose comments improved the quality of the paper. The majority of this work was performed while K.J.K.\ was at NASA GSFC where the work was supported through a NASA Earth and Space Science Fellowship. It was completed through funding from the Chief of Naval Research through the National Research Council. The numerical simulations were performed under a grant of High-End Computing resources to C.R.D.\ at NASA's Center for Climate Simulation.
}

\nocite{•}



\newpage

\newpage
\newpage

\newpage

\end{document}